\documentclass[noacm, acmsmall]{acmart}
\usepackage{amsmath}
\usepackage[ruled, vlined, linesnumbered]{algorithm2e}
\usepackage[utf8]{inputenc}
\usepackage{xspace}
\usepackage{enumitem}
\usepackage{bm}
\usepackage{amsthm}
\usepackage{xcolor}
\usepackage{svg}

\usepackage{xcolor}

\newcommand{\name}{Yannakakis$^+$\xspace}

\newcommand{\Q}{\mathcal{Q}}

\newcommand{\C}{\mathcal{C}}

\newcommand{\revise}[1]{{\color{black}#1}}

\newcommand{\J}{\mathcal{J}}
\newcommand{\T}{\mathcal{T}}
\newcommand{\A}{\mathcal{A}}

\renewcommand{\O}{\mathcal{O}}

\usepackage{stmaryrd}

\newcommand{\cbn}[1]{{\color{black}#1}}

\usepackage{mathtools}

\SetKw{KwAnd}{and}
\SetKw{KwOr}{or}

\usepackage[english]{babel}
\usepackage{amsthm}
\newtheorem{theorem}{Theorem}[section]
\newtheorem{lemma}[theorem]{Lemma}

\usepackage{listings}
\usepackage{color}
\definecolor{dkgreen}{rgb}{0,0.6,0}
\definecolor{gray}{rgb}{0.5,0.5,0.5}
\definecolor{mauve}{rgb}{0.58,0,0.82}
\newcommand{\IN}{N}
\newcommand{\OUT}{M}

\usepackage{tikz}

\usepackage[ruled, vlined, linesnumbered]{algorithm2e}
\SetKw{KwOr}{or}
\SetKw{KwAnd}{and}
\SetKwFor{ParFor}{for}{do in parallel}{end}

\usepackage{colortbl}

\DeclareMathOperator*{\moplus}{\text{\raisebox{0.25ex}{\scalebox{0.7}{$\bigoplus$}}}}
\DeclareMathOperator*{\motimes}{\text{\raisebox{0.25ex}{\scalebox{0.7}{$\bigotimes$}}}}

\usepackage{adjustbox}
\usepackage{graphicx}  
\usepackage{caption}
\usepackage{subfigure}
\usepackage{mdframed}
\usepackage{amsmath}
\usepackage{lineno}
\usepackage{placeins}
\usepackage{multirow}
\usepackage{multicol}
\usepackage{longtable}
\usepackage{tabularx}
\usepackage{colortbl}
\usepackage{xcolor}
\usepackage{hyperref} 
\usepackage{caption}  
\usepackage{forest}

\usepackage{listings}

\usetikzlibrary{arrows.meta, positioning}

\usepackage{ifthen}
\newboolean{techreport}
\setboolean{techreport}{true}

\title{\name: Practical Acyclic Query Evaluation with Theoretical Guarantees}



\renewcommand\footnotetextcopyrightpermission[1]{}

\author{Qichen Wang}
\authornote{Both authors contributed equally to this research.}
\email{qcwang@comp.hkbu.edu.hk}
\affiliation{%
  \institution{Hong Kong Baptist University}
  \country{Hong Kong SAR}
}

\author{Bingnan Chen}
\authornotemark[1]
\email{bchenba@cse.ust.hk}
\affiliation{%
  \institution{Hong Kong University of Science and Technology}
  \country{Hong Kong SAR}
}

\author{Binyang Dai}
\email{bdaiab@ust.hk}
\affiliation{%
  \institution{Hong Kong University of Science and Technology}
  \country{Hong Kong SAR}
  }

\author{Ke Yi}
\email{yike@cse.ust.hk}
\affiliation{%
  \institution{Hong Kong University of Science and Technology}
  \country{Hong Kong SAR}
}

\author{Feifei Li}
\email{lifeifei@alibaba-inc.com}
\affiliation{%
 \institution{Alibaba Group}
 \country{China}
}

\author{Liang Lin}
\email{yibo.ll@alibaba-inc.com}
\affiliation{%
  \institution{Alibaba Group}
  \country{China}
}

\begin{abstract}
  Acyclic conjunctive queries form the backbone of most analytical workloads, and have been extensively studied in the literature from both theoretical and practical angles.  However, there is still a large divide between theory and practice.  While the 40-year-old Yannakakis algorithm has strong theoretical running time guarantees, it has not been adopted in real systems due to its high hidden constant factor. In this paper, we strive to close this gap by proposing \name, an improved version of the Yannakakis algorithm, which is more practically efficient while preserving its theoretical guarantees.  Our experiments demonstrate that \name consistently outperforms the original Yannakakis algorithm by 2x to 5x across a wide range of queries and datasets.

  Another nice feature of our new algorithm is that it generates a traditional DAG query plan consisting of standard relational operators, allowing \name to be easily plugged into any standard SQL engine.  Our system prototype currently supports four different SQL engines (DuckDB, PostgreSQL, SparkSQL, and AnalyticDB from Alibaba Cloud), and our experiments show that \name is able to deliver better performance than their native query plans on \cbn{160} out of the \cbn{162} queries tested, with an average speedup of 2.41x and a maximum speedup of 47,059x.
\end{abstract}

\begin{document}

\maketitle
\begin{acks}
This work was supported by Hong Kong RGC Grants (Project No. 12200524, 16205422, 16204223, 16203924, C2004-21GF, and C2003-23Y) and an AIR grant from Alibaba Cloud.
\end{acks}
\sloppy
\section{Introduction}
\label{sec:intro}
Selection-join-projection-aggregation queries, a.k.a. \textit{conjunctive queries (CQs)}, form the backbone of most analytical workloads\footnote{\revise{The traditional definition of conjunctive queries does not consider aggregations.  The incorporation of aggregations is introduced in \cite{Abo2016FAQ, Joglekar2016AJAR}  under the semiring framework; see Section~\ref{sec:defCQ} for details. }}.  The following query, which is a slightly simplified version of TPC-H Query 9 \cite{tpch}, is one such example:
\begin{lstlisting}
SELECT n_name, o_orderkey, l_returnflag, SUM(ps_supplycost * l_quantity) AS part_cost
FROM nation, supplier, part, orders, lineitem, partsupp
WHERE o_orderdate < DATE '1996-12-31' and o_orderdate > 
  DATE '1996-01-01' and p_name LIKE '%blue%' 
  and o_orderkey = l_orderkey and ps_suppkey = l_suppkey 
  and ps_partkey = l_partkey and p_partkey = l_partkey 
  and s_suppkey = l_suppkey and s_nationkey = n_nationkey
GROUP BY n_name, o_orderkey, l_returnflag;
\end{lstlisting}

Due to their central importance, how to evaluate conjunctive queries efficiently has been extensively studied in the database community, from both practical and theoretical angles.  The predominant approach, implemented in most relational engines, aims to find an optimal query plan that takes the form of a directed acyclic graph (DAG).  The leaves of the DAG correspond to the input relations, while each internal node represents a relational operator, which can be either unary (selection, projection, and aggregation) or binary (join and semi-join), and the root node of the DAG yields the  query results.  

\begin{example}
\label{ex:Q9DuckDB}
We ran the query above in DuckDB, a popular column-based relational engine especially optimized for analytical workloads. The query plan it used is as follows (we rename the join attributes and use the natural join syntax):
\[
\begin{aligned}
    (1) & \quad J_1 \gets \pi_{\text{partkey}, \text{p\_name}}\sigma_{\text{p\_name } \mathtt{ LIKE } \dots}\left(\mathtt{part}\right)\Join \pi_{\text{partkey,orderkey,suppkey,l\_returnflag,l\_quantity}} \left(\mathtt{lineitem}\right); \\ 
    (2) & \quad J_2 \gets \pi_{\text{orderkey}}\sigma_{\cdots < \text{o\_orderdate} < \cdots}(\mathtt{orders}) \Join J_1; \\
    (3) & \quad J_3 \gets \pi_{\text{suppkey}, \text{nationkey}} \text{supplier} \Join \pi_{\text{n\_name}, \text{nationkey}} \mathtt{nation}; \\
    (4) & \quad J_4 \gets J_3 \Join J_2; \\
    (5) & \quad J_5 \gets J_4 \Join \pi_{\text{partkey}, \text{suppkey}, \text{ps\_supplycost}}\left( \mathtt{partsupp}\right) ;\\
    (6) & \quad Q\gets \gamma_{\text{n\_name,orderkey,l\_returnflag},\mathtt{SUM}(\text{ps\_supplycost*l\_quantity})}(J_5). 
\end{aligned}
\]
\end{example}

However, from the theoretical angle, this query plan is sub-optimal for the following two reasons: First, there might be many \textit{dangling tuples} that are unnecessarily involved in the joins, especially when the query has some highly selective predicates.   For instance, the predicate on $\mathtt{o\_orderdate}$ may filter out a large portion of the $\mathtt{orders}$ table, which means that many intermediate join results in $J_1$ will not be able to join with any tuple in the $\mathtt{orders}$ table, hence become \textit{dangling}. These dangling tuples could blow up the intermediate join size to $O(N^\rho)$ in the worst case, where $N$ is the input size and $\rho$ is the \textit{fractional edge cover number}\footnote{$\rho=4$ for TPC-H Q9.  Please see \cite{AGM} for the precise definition of $\rho$, but this is not crucial for understanding the paper.} of the query.  Second, this plan evaluates the full multi-way join before the aggregation.  This can be sub-optimal since the full join size (denoted by $F$ subsequently) can be much larger than the final query output size (denoted by $M$), which is equal to the number of groups.  In particular, when the aggregation does not have a $\mathtt{GROUP\ BY}$ clause, \revise{it aggregates all join results into a single value, hence $M=1$}.

In practice, nevertheless, these two potential risks may not materialize because data is often ``nice'': We ran the query on the TPC-H benchmark dataset with a scale factor (SF) of 500 in DuckDB, and it finished in just 9.2 seconds. In particular, this is because all the joins in this query are between a primary key (PK) and a foreign key (FK), which limits all intermediate join sizes, as well as the full join size $F$, to at most the input size $N$.  To expose the risk, we removed the PK constraints and duplicated an SF-100 dataset 5 times.  This results in a dataset of the same size, but each PK now has 5 copies.  This turns the joins into many-to-many joins, and the intermediate join sizes are no longer bounded by $N$.  On this dataset, DuckDB's running time blows up to 488 seconds, a 50x increase.  We have also tested other benchmarks with naturally occurring many-to-many joins, such as LSQB \cite{LSQB} and JOB \cite{job}, and observed similar phenomenon (please see Section \ref{sec:exp} for detailed results).  

Back to the theory side, there is actually a 40-year-old solution that already addressed these issues when the query is \textit{acyclic} (TPC-H Q9 is acyclic, and the formal definition will be given in Section \ref{sec:prelimiary}). In 1981, Yannakakis \cite{yannakakis1981algorithms} gave an algorithm that has a worst-case running time of $O(N+\OUT)$ or $O(\min(N\OUT, F))$, depending on whether the query has a certain property known as \textit{free-connex} (detailed definition given in Section \ref{sec:prelimiary}).  Note that such running times are especially appealing when $M$ is small, which is often the case for analytical queries that return aggregated results.  Furthermore, the $O(N+M)$ time, which is achievable for free-connex queries, is clearly asymptotically optimal.
Yannakakis' algorithm achieves these running times based on two key ideas: (1) use a series of semi-joins to remove all the dangling tuples before doing any joins, and (2) push the aggregations over joins as much as possible.  

Unfortunately, despite its nice theoretical guarantees, Yannakakis' algorithm has not been adopted in any query engines due to its large hidden constant factor \cite{ThomasPODS24}.  Indeed, we tested Yannakakis' algorithm in DuckDB on the TPC-H dataset with SF=500, and it took 21.3 seconds to evaluate Query 9, more than double that of DuckDB's query plan shown in Example \ref{ex:Q9DuckDB}.  Similar results have also been observed in \cite{gottlob2023structure}.  On the 5-copy dataset, however, we do see a significant improvement: Yannakakis' algorithm still runs in around 21 seconds (thanks to its worst-case guarantee), much faster than DuckDB's query plan which took 488 seconds.  



\subsection{Our Contributions}

This paper presents \name, an improved version of the Yannakakis algorithm, with the following properties:

\begin{enumerate}[leftmargin=*]
    \item It enjoys the same theoretical guarantee as the original Yannakakis algorithm on acyclic queries, i.e., it runs in $O(N+M)$ time if the query is free-connex, and $O(\min(NM,F))$ time otherwise. 
    \item It is more practically efficient than the Yannakakis algorithm on both PK-FK joins and many-to-many joins.  It consistently outperforms the Yannakakis algorithm by 2x to 5x (the maximum speedup is 87x) across four different SQL engines and a variety of queries/datasets.  It thus covers the shortcomings of the Yannakakis algorithm on PK-FK joins, while extending its gain on many-to-many joins, as well as on queries that involve both types of joins.  This makes \name the method of choice for a wide range of queries and datasets: Out of a total of 162 queries tested, \name is able to improve the SQL engines' own plans on 160 of them, with an average speedup of 2.41x and a maximum speedup of 47,059x.
    \item \name is also pure relational, in the sense that it can be formulated as a DAG query plan consisting of standard relational operators (see Table \ref{tbl:operators} for the operators that are needed).  In fact, we were able to implement \name completely outside a SQL engine, by generating the query plan in the form of SQL statements.  This allows \name to be used as a simple plug-in on top of any SQL engine, modulo minor changes in the syntax of the generated SQLs.  
\end{enumerate}

Furthermore, as many other queries can be reduced to acyclic CQs, such as cyclic CQs, queries with conjunctive sub-queries, unions and differences of CQs, top-$k$ queries, etc, \name can also be used to improve their evaluation by combining with other techniques. 
We describe these extensions in Section \ref{sec:general}.

\paragraph{Technical highlights}
The practical improvements from Yannakakis to \name are mostly driven by the following two observations.  First, the original Yannakakis algorithm, due to its theoretical motivation, separates the evaluation process into two distinct stages: The first stage uses two passes of semi-joins to remove all the dangling tuples, which takes $O(N)$ time.  Then the second stage uses a series of aggregation-joins to compute the query results, which takes $O(M)$ time (assuming the query is free-connex).  While theoretically clean, this separation incurs unnecessary computational overheads.  In \name, we push some aggregation-joins to before the semi-joins as much as possible, which is important since the aggregations can greatly reduce the data size, especially for queries with a small query output size $M$, while each join can remove a relation.  Furthermore, we also reduce the number of semi-joins needed; in particular, for a class of queries known as \textit{relation-dominated}, no semi-join is used at all.  A possibly undesirable consequence of removing some of the semi-joins is that not all dangling tuples are removed, so a technical challenge in our development is to prove that the remaining dangling tuples do not affect the worst-case running time. In Section \ref{sec:acyclic}, we describe these changes that we make to the Yannakakis algorithm.

Second, both Yannakakis and \name actually generate a family of query plans instead of a single one.  Theoretically, all these plans have the same asymptotic running time, but they differ in the hidden constant.  Thus, it is important to pick an optimal (or near-optimal) plan from this family.  Towards this end, we design a query optimizer tailored for \name.  Our optimizer follows the standard query optimization pipeline, consisting of a rule-based component and a cost-based component.  However, we must introduce some changes to both components, since \name has a different search space of query plans that existing query optimization methods do not cover.  We describe our \name optimizer in Section \ref{sec:opt}.

\begin{table}[t]
\begin{adjustbox}{max width= 0.6\textwidth}
\begin{tabular}{clccc}
\hline
Operator                     & SQL Query                                            & Complexity                        \\ \hline
Selection($\sigma_{f}(R)$)   &  \begin{lstlisting}[mathescape=true] 
SELECT * FROM R WHERE f; 
 \end{lstlisting}     &   $O(|R|)$                \\ \hline
Projection($\pi_{\bm{E}}(R)$) & \begin{lstlisting}[mathescape=true]
SELECT $\bm{E}$, $\oplus(v)$ AS $v$
FROM $R$ GROUP BY $\bm{E}$;
\end{lstlisting} & $O(|R|)$ \\ \hline
Join($R_1 \Join R_2$)     &   \begin{lstlisting}[mathescape=true]
SELECT *, $R_1.v\otimes R_2.v$ AS $v$ 
FROM $R_1$ NATURAL JOIN $R_2$;
\end{lstlisting} &  $\begin{aligned} &O(|R_1|+|R_2| \\&+|R_1\Join R_2|) \end{aligned}$                                      \\ \hline
SemiJoin($R_1 \ltimes R_2)$    & \begin{lstlisting}[mathescape=true]
SELECT * FROM $R_1$ WHERE 
$R_1.key$ in (SELECT 
DISTINCT $R_2.key$ FROM $R_2$);
\end{lstlisting}
&  $O(|R_1|+|R_2|)$          \\ \hline
\end{tabular}
\end{adjustbox}
\caption{Summary of relation operators and the corresponding SQL queries, where $v$ represents the annotation}
\label{tbl:operators}
\end{table}

\subsection{Related Work}
\label{sec:related}
Efficient evaluation of conjunctive queries has been extensively studied in the literature.   \revise{Sideways information passing (SIP) \cite{abiteboul1995foundations} is a widely used technique for query optimization that reduces intermediate results, and it is adopted by systems such as DBMS X and Amazon Redshift \cite{gupta2015amazon}. However, unlike the Yannakakis algorithm, SIP does not remove all dangling tuples when the query contains more than two relations, which can lead to suboptimal plans. Meanwhile, worst-case optimal join algorithms (WCOJ) \cite{Ngo2018worst} perform better on highly cyclic queries.}  The Yannakakis algorithm and WCOJ can be combined using the generalized hypertree decomposition framework \cite{gottlob99Hypertree, gottlob2009generalized} to provide running times that depend on the level of cyclicity of the query, measured by various \textit{width parameters} \cite{gottlob99Hypertree, gottlob2009generalized,grohe2014constraint,Abo2017Shannon}.  \revise{
Recent studies extend the Yannakakis algorithm to support different operators and scenarios, including projections \cite{bagan2007acyclic}, aggregations \cite{Abo2016FAQ, Joglekar2016AJAR}, unions \cite{Nofar2021Union}, differences \cite{Hu2023Computing}, comparisons \cite{wang2022conjunctive}, top-k queries \cite{WANG2024Relational}, dynamic \cite{idris2017dynamic, wang2023change} or secure \cite{wang2021secure} query processing. While these developments are promising, very few of them have made their way to real systems yet. }

\revise{Several recent works \cite{DBLP:conf/cidr/YangZYK24, Robust2024Birler, bekkers2024instance, zhao2025debunkingmythjoinordering} focus on implementing the Yannakakis algorithm efficiently within a particular database engine, but they do not change the algorithm itself. In contrast, we have improved the algorithm. Thus, their techniques are complementary to ours and can be combined with our approach when \name is integrated into their target database engine. Furthermore, \name is aimed at conjunctive queries with (group-by) aggregations, while \cite{DBLP:conf/cidr/YangZYK24, Robust2024Birler, bekkers2024instance} only considers full joins. For cyclic queries, RelationalAI \cite{RelationalAI} and Umbra \cite{adopting2020Freitag} adopt WCOJ inside databases, but they have to build the engine from ground up, since WCOJ does not directly generate a DAG plan using standard operators available in existing systems.
Our method can also be combined with WCOJ to handle cyclic queries, as explained in Section~\ref{sec:general}. We have not implemented this combination, since we prefer a purely relational approach that yields standard DAG query plans. }



We prove the worst-case running time of \name based on the running times in Table \ref{tbl:operators}.  If indexes are available, some of these operators can be executed faster, e.g., selection can be sped up to $O(\log |R| + |\sigma_f(R)|)$ when there is a B-tree index and $f$ is a range predicate; assuming $|R_2|>|R_1|$ and there is a hash-index on $R_2$, then the join can be computed in time $O(|R_1| + |R_1 \Join R_2|)$.  There is an extensive literature on indexing techniques \cite{hentschel2018column,ding2019ai,chaudhuri1997efficient,lang2019performance}.  The availability of indexes can only make \name run faster, so all our theoretical guarantees are not affected; in practice, it can be factored into our cost-based optimizer to pick the best plan in the \name family.  


Cost-based optimization is an important step in reducing the hidden constant factor of query plans, which is also used in \name. It involves three main components: cardinality estimation (CE), cost model (CM), and plan enumeration (PE). CE employs data statistics and assumptions on data distribution to estimate tuple counts using synopsis-based (e.g., histogram-based \cite{his-based1, his-based2} and sketch-based \cite{sketch-based1, sketch-based2}), sampling-based \cite{sample-based1, sample-based2, sample-based3, sample-based4}, and learning-based methods \cite{learn-based1, learn-based2, learn-based3}. CM translates the database state (which relations are in memory, availability of indexes, etc.) and cardinality estimates into execution costs, with traditional models defined by experts and modern, adaptive learning-based methods \cite{cm-1, cm-2, cm-3}. PE identifies the query plan with minimal cost, employing both non-learning (dynamic programming \cite{pe-1, pe-2, pe-3}, top-down strategies \cite{pe-4, pe-5}) and learning-based approaches \cite{pe-6, pe-7}.  For CE and CM, we can use existing techniques.  However, we have to design new PE methods, since \name has a different search space.

\section{Preliminaries}
\label{sec:prelimiary}
\subsection{Conjunctive Queries}
\label{sec:defCQ}
We consider \textit{conjunctive queries (CQs)} of the following form:
\begin{equation}
\label{eq:def}
    \Q = \pi_{\O} \left( R_1(\A_1) \Join R_2(\A_2) \Join \cdots \Join  R_n(\A_n) \right),
\end{equation}
where each $R_i(\A_i)$ is a relation with a set of attributes $\A_i$, for $i=1,2\dots,n$.  The same relation may appear more than once with attribute renamings (i.e., self-joins);  we consider them as logical copies of the same relation.  We use $\bm{R} =\{R_1, \cdots, R_n\}$ to denote the set of all relations in the query, and 
$\A=\A_1 \cup \A_2 \cup \cdots \cup \A_n$ the set of all attributes.   For a subset of the relations $\mathcal{S} \subseteq \bm{R}$, let $\A(\mathcal{S})$ be the attributes that appear in $\mathcal{S}$; and define $\bar{\A}_i = \A(\bm{R} - \{R_i\})$, i.e., all attributes except those that only appear in $R_i$.


The original work of Yannakakis only considered a (distinct) projection $\pi_\O$ after the multi-way join.  The extension to aggregations is made in \cite{Abo2016FAQ,Joglekar2016AJAR}, who use semirings to formalize the types of aggregations that can be supported.  Let $(\mathbb{S}, \oplus, \otimes)$ be a communicative semiring, where $\mathbb{S}$ is the ground set, with $\oplus$ and $\otimes$ being its ``addition'' and ``multiplication'', respectively.  
Each input tuple $t \in R_i$ is associated with an \textit{annotation} $v_i(t) \in \mathbb{S}$.  These annotations are propagated through the join and projection, as follows.  The annotation for any tuple $t$ in the join results $\mathcal{J}= R_1(\A_1) \Join R_2(\A_2) \Join \cdots \Join  R_n(\A_n)$ is the $\otimes$-aggregate of all the tuples, one from each relation, that make up $t$:
\[
    v(t) := \motimes_{R_i(\A_i) \in \Q} v_i (\pi_{\A_i} t).
\]
Then $\pi_{\O}$ performs a $\oplus$-aggregation grouped by  $\O$, i.e., the annotation of each tuple $t$ in the final query results is 
\[
    v(t) := \moplus_{\forall t' \in \mathcal{J}, \pi_{\O} t' = t} v'(t').
\]
The attributes in $\O$ are called the \textit{output attributes}. Specially, if $\O = \emptyset$ and $\J\ne \emptyset$, then the query returns the empty tuple $\langle \rangle$ associated with an annotation that  aggregates all the join results:
\[
    v(\langle \rangle) = \moplus_{\forall t' \in \mathcal{J}} v'(t').
\]
If $\O = \A$ (i.e., $\Q = \mathcal{J}$), then the query is called a \textit{full query}, which does not perform any $\oplus$-aggregation.
  
Such a conjunctive query with properly defined annotations is  equivalent to the following SQL query:
\begin{lstlisting}[mathescape=true]
 SELECT      $\O$, $\oplus(v_1 \otimes \cdots \otimes v_n)$
 FROM        $R_1$ NATURAL JOIN $\cdots$ NATURAL JOIN $R_n$
 GROUP BY    $\O$;
\end{lstlisting}

\begin{example}
\label{ex:query}
     TPC-H Query 9 in Section \ref{sec:intro} can be represented as the following conjunctive query over the semiring $(\mathbb{R}, +, \cdot)$:
    \[
        \begin{aligned}
        \Q_1= & \pi_{x_1, x_2, x_8} \left((R_1(x_1, x_2, x_3, x_4) \Join R_2(x_2, x_5) \Join R_3(x_3, x_4) \right.\\ 
        &\left.\Join R_4(x_3, x_6) \Join R_5(x_4, x_7) \Join R_6(x_7, x_8)\right)
        \end{aligned}
    \]
    where $R_1, R_2, R_3, R_4, R_5, R_6$ correspond to the relations $\mathtt{lineitem}$, $\mathtt{orders}$, $\mathtt{partsupp}$, $\mathtt{part}$, $\mathtt{supplier}$, and $\mathtt{nation}$, respectively, while $x_1, x_2, x_3, x_4, x_5, x_6, x_7, x_8$ correspond to the attributes $\mathtt{l\_returnflag}$, $\mathtt{orderkey}$, $\mathtt{partkey}$, $\mathtt{supplierkey}$, $\mathtt{o\_orderdate}$, $\mathtt{p\_name}$, $\mathtt{nationkey}$, $\mathtt{n\_name}$.  Note that we have dropped unnecessary columns and renamed the join attributes to fit the natural join syntax.  
    For all tuples in $R_2, R_4, R_5, R_6$, their annotations are set to $1$. For each tuple $t \in R_3$, set $v(t) := \mathtt{ps\_supplycost}$; for each tuple $t \in R_1$, set $v(t) := \mathtt{l\_quantity}$.

    We have also omitted the selection operators $\sigma$, which can always be pushed down to the input relations.  They can be handled by a table scan or more efficiently by index retrieval if  available, which are  issues orthogonal to this work.
    \qed
\end{example}

Note that this semiring formulation unifies most cases of the aggregation operator $\gamma$ into $\pi$.  In particular, projection can be considered as a special case of aggregation on the boolean semiring $(\{\textsf{False}, \textsf{True}\}, \land, \lor)$, and all tuples in the database are assigned annotation $\textsf{True}$.  By choosing the semiring and annotations appropriately, this formulation incorporates a variety of aggregation queries.  For example, in addition to the commonly used $(\mathbb{R}, +, \cdot)$ in the above example, the semiring $(\mathbb{R}, \mathtt{MAX}, +)$ allows us to compute  aggregations like $\mathtt{MAX(ps\_availqty - l\_quantity)}$ by setting the annotations in $R_3$ to $\mathtt{ps\_availqty}$ and the annotations in $R_1$ to $-\mathtt{l\_quantity}$.


We will also make use of some other standard relational operators including selection, union, semi-join, and order-by.  These operators will not change the annotations of the tuples.  In particular, a semi-join $R_1 \ltimes R_2$ returns all the tuples in $R_1$ that can join with at least one tuple in $R_2$.  Here, the tuples in $R_1$ retain their annotations in $R_1$ after the semi-join, and the annotation in $R_2$ are irrelevant. 

When analyzing the running time of an algorithm, we adopt the standard RAM model of computation and consider the \textit{data complexity}, i.e., the query size is taken as a constant.  We will measure the running time of a query evaluation algorithm using three parameters: the total input size $\IN = \sum_{i=1}^n |R_i|$, the query output size $\OUT$, and the full join size $F = |R_1(\A_1) \Join \cdots \Join R_n(\A_n)|$.  Note that $M=F$ for a full query, but $M$ could be much smaller than $F$ for a non-full query.  

\subsection{Classification of CQs}
In the study of CQs, the following classes have been identified to bear complexity-theoretical significance:  

\noindent\textbf{Acyclic CQs.} There are many equivalent definitions for acyclic CQs, and we adopt the one based on \emph{join trees} \cite{beeri1983desirability, fagin1983degrees}.  A CQ $\Q$ is acyclic if there exists a tree $\T$ satisfying the following properties: (1) the set of nodes in $\T$ have a one-to-one mapping to the set of relations in $\Q$; and (2) for each attribute $x \in \A$, all nodes of $\T$ containing $x$ form a connected subtree of $\T$.  The join tree may not be unique; the  GYO algorithm  \cite{university1980universal,yu1979algorithm} can be used to decide whether a given query $\Q$ is acyclic, and if yes, find all possible join trees for $\Q$.  Because of the one-to-one mapping between the relations and the tree nodes, we may use these two terms interchangeably on a fix join tree $\T$.  For a node/relation $R_i(\A_i)$ in $\T$, we often use $R_p(\A_p)$ to denote its parent node, and $\C_i$ its children.  

Note that the acyclicity of a query does not concern its output attributes, which are instead considered in the following sub-classes of acyclic CQs. 

\noindent\textbf{Free-connex CQs.} 
A CQ $\Q$ is \emph{free-connex} if both $\Q$ and $\Q \Join [\O]$ are acyclic, where $[\O]$ denotes a relation with all output attributes \cite{bagan2007acyclic}.  This definition, however, is not easy to use in query evaluation, since $\Q$ and  $\Q \Join [\O]$ have different join trees.   
In this paper, we use the following equivalent definition\footnote{\ifthenelse{\boolean{techreport}}{All proofs are given in the appendix.}{Due to space constraints, all proofs are given in the technical report \cite{techreport}.}} that uses a single join tree.
\begin{lemma}
\label{lem:free-connex}
    A CQ $\Q$ is \emph{free-connex} if and only if it has a join tree $\T$ with a subtree $\T_n$ containing the root node that satisfies two conditions: (1) $\O \subseteq \A(\T_n)$, where $\A(\T_n)$ represents the set of all attributes present in $\T_n$, and (2) for any non-root node $R(\A) \in \T_n$ with parent $R_p(\A_p)$, $\A \cap \A_p \subseteq \O$. 
Such a $\T$ is called a free-connex join tree of $\Q$, and $\T_n$ is referred to as its \textit{connex subset}.
\end{lemma}

In addition, we identify another sub-class of queries:

\noindent\textbf{Relation-dominated CQs.} A CQ $\Q$ is \emph{relation-dominated} if $\Q$ is acyclic and there exists a relation $R_i(\A_i)$ such that $\O \subseteq \A_i$.  We call $R_i(\A_i)$ the \textit{dominating relation}, and the join tree with $R_i(\A_i)$ as the root the relation-dominated join tree of $\Q$.  Note that for the special case $\O = \emptyset$, the query is dominated by any of its relations, and any of its join trees is a relation-dominated join tree.



\begin{figure}[h]
    \centering
    \subfigure[Join tree $\T_1$ for $\Q_1$]
    {
        \begin{minipage}[t]{0.38\columnwidth}
        \centering
        \resizebox{\linewidth}{!}{
            \begin{tikzpicture}[
              edge from parent/.style={draw, thick}, 
              level 1/.style={sibling distance=13mm, level distance=10mm},
              level 2/.style={sibling distance=25mm, level distance=10mm},
              level 3/.style={sibling distance=10mm, level distance=10mm},
              every node/.style={draw, ellipse, minimum width=20mm, minimum height=10mm, font= \Huge, ultra thin}
              ]

              \node[fill={rgb,255:red,255; green,205; blue,210}, fill opacity=1, text opacity=1] (R5) at (0, 0) {$R_5(x_4, x_7)$}  
                child {node[fill={rgb,255:red,135; green,206; blue,250}, fill opacity=1, text opacity=1] (R1) at (-2.5, -0.7) {$R_1(\underline{x_1}, \underline{x_2}, x_3, x_4)$}
                  child {node[fill={rgb,255:red,255; green,138; blue,128}, fill opacity=1, text opacity=1] (R3) at (0, -0.7) {$R_3(x_3, x_4)$}
                    child {node[fill={rgb,255:red,225; green,190; blue,231}, fill opacity=1, text opacity=1] (R4) at (0, -0.7) {$R_4(x_3, x_6)$}}
                  }
                  child {node[fill={rgb,255:red,80; green,255; blue,80}, fill opacity=0.8, text opacity=1] (R2) at (2, -0.7) {$R_2(\underline{x_2}, x_5)$}}
                }
                child {node[fill={rgb,255:red,255; green,255; blue,102}, fill opacity=1, text opacity=1] (R6) at (2, -0.7) {$R_6(x_7, \underline{x_8})$}}
                ;
            \end{tikzpicture}
        }
        \label{fig:ex2_1_jt_1}
        \end{minipage}
    }
    \subfigure[Free-connex join tree $\T_2$ for $\Q_2, \Q_3$]
    {
        \begin{minipage}[t]{0.51\linewidth}
        \centering
        \resizebox{\linewidth}{!}{
            \begin{tikzpicture}[
              edge from parent/.style={draw, thick}, 
              level 1/.style={sibling distance=13mm, level distance=10mm},
              level 2/.style={sibling distance=13mm, level distance=10mm},
              level 3/.style={sibling distance=13mm, level distance=10mm},
              every node/.style={draw, ellipse, minimum width=20mm, minimum height=11mm, font=\Huge, thick}
              ]
              \filldraw[fill=lightgray!80, opacity=0.5, rounded corners=10pt, draw=none] 
                (2.5, -0.5) ellipse (6 and 2.35); 

              \node[font=\Huge, text=darkgray, draw=none] at (5.4, 0.5) {Connex};
              \node[font=\Huge, text=darkgray, draw=none] at (5.4, -0.1) {Subset};

              \node[fill={rgb,255:red,135; green,206; blue,250}, fill opacity=1, text opacity=1] (R1) at (0, 0) {$R_1(\underline{x_1}, \underline{x_2}, \underline{x_3}, x_4)$}
                child {node[fill={rgb,255:red,255; green,205; blue,210}, fill opacity=1, text opacity=1] (R5) at (-2.4, -2.3){$R_5(x_4, x_7)$}
                    child {node[fill={rgb,255:red,255; green,255; blue,102}, fill opacity=1, text opacity=1] (R6) at (0, -0.5){$R_6(x_7, x_8)$}}
                  }
                child {node[fill={rgb,255:red,255; green,138; blue,128}, fill opacity=1, text opacity=1] (R3) at (-4.2, -0.7) {$R_3(\underline{x_3}, x_4)$}}
                child {node[fill={rgb,255:red,225; green,190; blue,231}, fill opacity=1, text opacity=1] (R4) at (-0.2, -0.7) {$R_4(\underline{x_3}, \underline{x_6})$}}
                child {node[fill={rgb,255:red,80; green,255; blue,80}, fill opacity=0.8, text opacity=1] (R2) at (3.2, -0.7) {$R_2(\underline{x_2}, \underline{x_5})$}};

            \end{tikzpicture}
        }
        \label{fig:ex2_1_jt_2}
        \end{minipage}
    }
    \caption{Two possible join trees for $\Q_1$, $\Q_2$ and $\Q_3$. The output attributes are underlined.}
    \label{fig:binary_plan}
\end{figure}
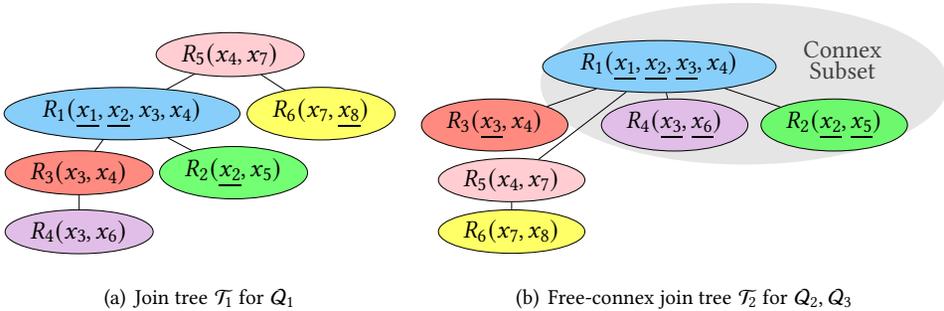

\begin{example}
\label{ex:freeconnex}
  TPC-H Query 9 ($\Q_1$ in Example~\ref{ex:query}) is an acyclic query with two possible join trees $\T_1$ and $\T_2$ shown in Figure~\ref{fig:ex2_1_jt_1} and Figure~\ref{fig:ex2_1_jt_2}. 
  
  $\Q_1$ is not free-connex.  But if we change the output attributes to $\O= \{x_1, x_2, x_3, x_5, x_6\}$, then the resulting query
  \[
    \Q_2 \gets \pi_{x_1, x_2, x_3, x_5, x_6} \left(\mathop{\Join}_{i \in [6]}R_i\right)
  \]
  is free-connex, with a free-connex join tree $\T_2$ shown in Figure~\ref{fig:ex2_1_jt_2}.  Note that $\T_1$ is not a valid free-connex join tree for $\Q_2$ because the join attributes between $R_1$ and $R_3$ contain a non-output attribute $x_4$.
 
Furthermore, if we change the output attributes to $\O= \{x_1\}$, then the query
  \[
    \Q_3 \gets \pi_{x_1} \left(\mathop{\Join}_{i \in [6]}R_i\right)
  \]
  is relation-dominated, by picking $R_1$ as the root of the join tree. \qed
\end{example}

\subsection{The Yannakakis Algorithm}
It is clear that all relation-dominated queries are free-connex queries, and all free-connex queries are acyclic queries.  Acyclic and free-connex queries are at the core in the theory of query evaluation: All acyclic queries can be evaluated in $O(\min(\IN \OUT, F))$ time~\cite{yannakakis1981algorithms}, while free-connex queries can be  evaluated in $O(\IN+\OUT)$ time~\cite{bagan2007acyclic}.  Both running times are achievable by the Yannakakis algorithm, which, on a given acyclic query $\Q$ with a join tree $\T$, works as follows:
\begin{enumerate}[leftmargin=*]
    \item Traverse the tree in the post-order; for each visited tree node $R_i$ and its parent node $R_p$, replace $R_p$ with $R_p \ltimes R_i$;
    \item Traverse the tree in the pre-order; for each visited non-leave node $R_i$, for each $R_c \in \C_i$, replace $R_c$ with $R_c \ltimes R_i$;
    \item Traverse the tree in the post-order again; for each visited tree node $R_i$, replace $R_p$ with $\left(\pi_{\A_p \cup \O} R_i\right) \Join R_p$ and remove $R_i$ from the tree.  
    \item Until only one node $R_r$ on the join tree, output $\pi_{\O} R_r$ as the query result. 
\end{enumerate}

\begin{example}
\label{ex:yan}
Using the join tree $\T_1$ in Figure \ref{fig:ex2_1_jt_1}, the Yannakakis algorithm yields the following query plan for $\Q_1$:

\begin{minipage}{0.9\linewidth}  
\setlength{\columnsep}{1pt}
\begin{multicols}{2}
\begin{enumerate}[leftmargin=*]
  \item[(1)] $R_1 \gets R_1 \ltimes R_2$; 
  \item[(3)] $R_1 \gets R_1 \ltimes R_3$;
  \item[(5)] $R_5 \gets R_5 \ltimes R_6$;
  \item[(7)] $R_1 \gets R_1 \ltimes R_5$; 
  \item[(9)] $R_4 \gets R_4 \ltimes R_3$;
  \item[(11)] $\mathcal{J}_1 \gets \pi_{x_2}R_2 \Join R_1$;
  \item[(13)] $\mathcal{J}_3 \gets \mathcal{J}_1 \Join \mathcal{J}_2$;
  \item[(15)] $\mathcal{J}_5 \gets \mathcal{J}_4 \Join R_6$;
\end{enumerate}
\columnbreak
\begin{enumerate}[leftmargin=*] 
  \item[(2)] $R_3 \gets R_3 \ltimes R_4$;
  \item[(4)] $R_5 \gets R_5 \ltimes R_1$;
  \item[(6)] $R_6 \gets R_6 \ltimes R_5$;
  \item[(8)] $R_3 \gets R_3 \ltimes R_1$;
  \item[(10)] $R_2 \gets R_2 \ltimes R_1$;
  \item[(12)] $\mathcal{J}_2 \gets \pi_{x_3}R_4 \Join R_3$;
  \item[(14)] $\mathcal{J}_4 \gets \pi_{x_1, x_2, x_4}\mathcal{J}_3 \Join R_5$;
  \item[(16)] $\Q_1 \gets \pi_{x_1, x_2, x_8} \mathcal{J}_5$. \qed
  \end{enumerate}
\end{multicols}  
\end{minipage}
\end{example}

  A linear complexity $O(\IN+\OUT)$ is clearly optimal; the optimality of the $O(\min(\IN \OUT, F))$ bound is still elusive, but it is nevertheless the best-known worst-case running time for acyclic but non-free-connex CQs. 

\section{\name}
\label{sec:acyclic}

In this section, we describe \name.  For now, we assume that a join tree $\T$ (a free-connex join tree for a free-connex CQ, and a relation-dominated join tree for a relation-dominated query) is given; we will discuss how to pick a good join tree in Section \ref{sec:opt}.

\subsection{First-round computation}
\label{sec:algo_yanna}




\name consists of two rounds.  The first round performs a post-order traversal on $\T$, as shown in Algorithm~\ref{alg:R1}.

\begin{algorithm}[h]
    \caption{First round post-order traversal}
    \label{alg:R1}
    \KwIn{A join tree $\T$ for the acyclic query $\Q$ on relations $\bm{R}$}
    \KwOut{A reduced join tree $\T'$ on relations $\bm{R}'$}
    Let $R_1, \cdots, R_n$ be arranged in some post-order of $\T$\;
    \ForEach{$i \in [n-1]$}{
        Let $R_p(\A_p)$ be the parent node of $R_i(\A_i)$ on $\T$\;
        \eIf {$R_i$ is a leaf node of $\T$ and $\A_i \cap \O \subseteq \A_p$}{
            $R_p \gets R_p \Join \left(\pi_{\A_p} R_i\right)$\;
            $\T \gets \T - \{R_i\}$, $\bm{R} \gets \bm{R} - \{R_i\}$\;
        } 
        {
            $R_i \gets \pi_{\O \cup \bar{\A}_i} R_i$\;
            $R_p \gets R_p \ltimes R_i$\;
        }
    }
    $R_n \gets \pi_{\O\cup \bar{\A}_n} R_n$\;
    \KwRet{$\T, \bm{R}$}
\end{algorithm}

Below, we use three examples to illustrate how the first round works.

\begin{example}
\label{ex:binary}
    First consider a simple query on two relations:  
    \[\begin{aligned}
        &\Q_4 \gets \pi_{x_1}\left(R_1(x_1, x_2) \Join R_2(x_2, x_3)\right).
    \end{aligned}\]
    Note that this query is relation-dominated, hence also free-connex, and the (only) free-connex join tree has $R_1$ as the root and $R_2$ as the leaf.
    
    The standard query plan used in most database systems for this query is 
\[
\begin{aligned}
    (1) &\quad \mathcal{J} \gets R_1 \Join R_2; & (2) & \quad \text{return }T \gets \pi_{x_1}\mathcal{J}. 
\end{aligned}
\]
    This plan takes $O(N+F)$ time; recall that $F=|R_1 \Join R_2|$ is the full join size.

    In contrast, the Yannakakis algorithm for this query achieves $O(N+M) = O(N)$ time \revise{(since $M \le N$ on this query)}, through the following plan:
\[
\begin{aligned}
    (1) &\quad R_1 \gets R_1 \ltimes R_2; & (2) & \quad R_2 \gets R_2 \ltimes R_1; \\ 
    (3) &\quad R_1 \gets R_1 \Join \pi_{x_2} R_2; & (4) & \quad \text{return } T \gets \pi_{x_1}R_1. \\ 
\end{aligned}
\]

Algorithm \ref{alg:R1} on this query yields the following plan:
\[
\begin{aligned}
    (1) &\quad R_1 \gets R_1 \Join \pi_{x_2} R_2; & (2) & \quad \text{return }T \gets \pi_{x_1} R_1. 
\end{aligned}
\]
Because only one relation remains after the first-round computation, \name terminates without needing to do the second round. We see that the \name plan is actually the same as the last two steps in the Yannakakis plan.  Essentially, the observation is that, for this query, the two semi-joins are not necessary; doing the last two steps directly, even with the presence of dangling tuples, still guarantees \revise{$O(N)$} time, which we will prove more formally and generally later.  

We ran the three query plans in DuckDB on the Epinion graph, where both $R_1$ and $R_2$ refer to the edge relation with 508,837 edges (namely, it is a self-join).  We use the $(\mathbb{N}, +, \cdot)$ smearing and set all input tuples' annotations to $1$, so the query returns the number of length-2 paths for each vertex $x_1$.  The standard plan took 0.507 s,  the Yannakakis plan took 0.243 s, while our new plan took 0.0366~s.
\qed
\end{example}

\begin{example}
\label{ex:fc}
    Next, consider $\Q_2$ from Example~\ref{ex:freeconnex}, which is free-connex but not relation-dominated.  This query has more than one free-connex join tree; in this example, we use $\T_2$ in Figure \ref{fig:ex2_1_jt_2}.  Then Algorithm \ref{alg:R1} yields the following steps:
\[
\begin{aligned}
    (1) &\quad R_5 \gets R_5 \Join \pi_{x_7} R_6; & (2) &\quad R_1 \gets R_1 \Join R_3; \\
    (3) &\quad R_1 \gets R_1 \Join \pi_{x_4} R_5;   & (4) & \quad R_1 \gets R_1 \ltimes R_2; \\
    (5) & \quad R_1 \gets R_1 \ltimes R_4;  & (6) &\quad R_1 \gets \pi_{x_1, x_2, x_3} R_1.
\end{aligned}
\]
The first three steps fall into the \textbf{if} part, since the output attributes in $R_3, R_5, R_6$ also appear in their parents.  We do early aggregation and join for these relations, which are then removed.  Steps (4)--(5) take the \textbf{else} part that does the semi-joins.  Note that line 8 in Algorithm \ref{alg:R1} is a no-op in this example.  Step (6) performs the final aggregation of line 10 in Algorithm \ref{alg:R1}.  We see Algorithm \ref{alg:R1} has reduced the query to a full join (which we will show is true for all free-connex queries):
\[
    \Q_2' \gets R_1(x_1, x_2, x_3) \Join R_2(x_2, x_5) \Join R_4(x_3, x_6),
\]
and the reduced join tree $\T_2'$ is shown in Figure \ref{fig:jointree2}. \qed
\begin{figure}[ht]
    \subfigure[Join tree $\T_1'$ for $\Q_1'$ after the first round.]
    {
    \centering
    \begin{minipage}[t]{0.47\columnwidth}
        \centering
        \resizebox{\linewidth}{!}{
            \begin{tikzpicture}[scale=0.8,
    every node/.style={draw, ellipse, minimum width=12mm, minimum height=6mm, font=\normalsize, ultra thin}
    ] 

    \begin{scope}
        \node[fill={rgb,255:red,255; green,205; blue,210}, fill opacity=1, text opacity=1] (R3) at (0,0) {$R_5(x_4, x_7)$};

        \node[fill={rgb,255:red,135; green,206; blue,250}, fill opacity=1, text opacity=1] (R4) at (-2.5,-1) {$R_1(\underline{x_1}, \underline{x_2}, x_4)$};

        \node[fill={rgb,255:red,255; green,255; blue,102}, fill opacity=1, text opacity=1] (R2) at (2.2,-1) {$R_6(x_7, \underline{x_8})$};

        \draw[-, line width=0.3mm] (R2) -- (R3); 
        \draw[-, line width=0.3mm] (R4) -- (R3); 
    \end{scope}
\end{tikzpicture}
        }
        \label{fig:jointree1}
    \vspace*{-0.2cm}
    \end{minipage}
    }
    \subfigure[Join tree $\T_2'$ for $\Q_2'$ after the first round.]
    {
    \begin{minipage}[t]{0.47\columnwidth}
        \centering
        \raisebox{0cm}{
            \resizebox{\linewidth}{!}{
                \begin{tikzpicture}[scale=0.8, every node/.style={draw, ellipse, minimum width=12mm, minimum height=6mm, font= \normalsize, ultra thin}] 
    
    \begin{scope}
        \node[fill={rgb,255:red,80; green,255; blue,80}, fill opacity=0.8, text opacity=1] (R4) at (-2.5,-1) {$R_2(\underline{x_2}, \underline{x_5})$};
        
        \node[fill={rgb,255:red,135; green,206; blue,250}, fill opacity=1, text opacity=1] (R3) at (0,0) {$R_1(\underline{x_1}, \underline{x_2}, \underline{x_3})$};
        
        \node[fill={rgb,255:red,225; green,190; blue,231}, fill opacity=1, text opacity=1] (R2) at (2.5,-1) {$R_4(\underline{x_3}, \underline{x_6})$};
            
        \draw[-, line width=0.3mm] (R2) -- (R3);

        \draw[-, line width=0.3mm] (R4) -- (R3);

    \end{scope}
\end{tikzpicture}
            }
        }
        \label{fig:jointree2}
    \vspace*{-0.2cm}
    \end{minipage}
    }
    \vspace*{-0.2cm}
    \caption{Two Jointrees for $\Q_1'$ and $\Q_2'$. }
    \label{fig:jointree12}
\end{figure}
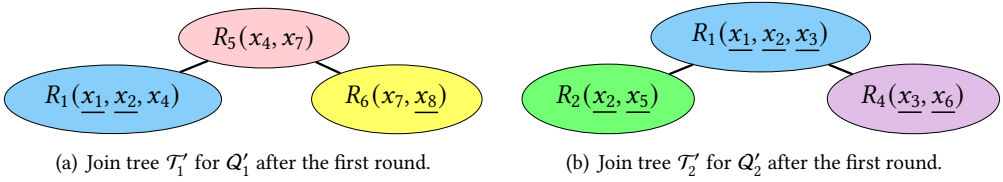
\end{example}

\begin{example}
\label{ex:Q3}
Finally, consider a non-free-connex query, $\Q_1$ from Example~\ref{ex:query}.  Suppose we use the join tree $\T_1$ in Figure~\ref{fig:ex2_1_jt_1}.  Then Algorithm \ref{alg:R1} yields the following query plan:
\[
\begin{aligned}
   (1) & \quad R_1 \gets R_1 \Join \pi_{x_2} R_2; & (2) & \quad R_3 \gets R_3 \Join \pi_{x_3} R_4; \\ 
    (3) &\quad R_1 \gets R_1 \Join R_3; & (4) &\quad R_1 \gets \pi_{x_1, x_2, x_4} R_1; \\
    (5) &\quad R_5 \gets R_5 \ltimes R_1; & (6) &\quad R_5 \gets R_5 \ltimes R_6; 
\end{aligned}
\]
and the reduced query is 
\[
    \Q_1' \gets \pi_{x_1, x_2, x_8} R_1(x_1, x_2, x_4) \Join R_5(x_4, x_7) \Join R_6(x_7, x_8)
\]
with the join tree $\T_1'$ in Figure \ref{fig:jointree1}.

Compared with the free-connex $\Q_2$, some non-output attributes of $\Q_1$ remain, but Algorithm \ref{alg:R1} did the best it can: The remaining attributes are either output attributes or join attributes (e.g., $x_4$ and $x_7$) that are ``shielded'' by the output attributes from below.
\qed
\end{example}

\paragraph{Correctness}
We will prove that the reduced query after the first round is equivalent to the original query.
\begin{lemma}
\label{lem:3.6}
    On any acyclic query $\Q$ and its join tree $\T$, Algorithm \ref{alg:R1} produces a query $\Q'$ that is equivalent to $\Q$.
\end{lemma}

\paragraph{Running time}
We see that all the operators in the first-round computation has running time $O(N)$, and none of them increases the data size.  This is clearly the case for all semi-joins and aggregations. Join is the only operator that may take more than linear time and enlarge the data size, but the join done in line 5 of Algorithm \ref{alg:R1} is between $R_p$ and $\pi_{\A_p}R_i$, and the latter's attribute set is a subset of the former.  This is thus essentially a semi-join if annotations are not concerned.  Since we are removing $R_i$, we need to use a join here to make sure that the annotations in $\pi_{\A_p}R_i$ are correctly multiplied by those in $R_p$. 

\begin{lemma}
\label{lem:firstround}
    The worst-case running time of Algorithm \ref{alg:R1} is $O(N)$.
\end{lemma}

\paragraph{Properties of the reduced query}
In addition to being equivalent to the original query, we can prove the following properties of the reduced query $\Q'$, which will be useful in the second round:

\begin{lemma}
\label{lem: output-only}
For a given acyclic query $\Q$, Algorithm \ref{alg:R1} returns a reduced query $\Q'$ that only has 
\begin{enumerate}[leftmargin=*]
    \item output attributes and join attributes;
    \item output attributes (i.e., $\Q'$ is a full query) if $\Q$ is free-connex;
    \item one relation consisting of only output attributes if $\Q$ is relation-dominated. 
\end{enumerate}
\end{lemma}


Combining Lemma~\ref{lem:firstround} and Lemma~\ref{lem: output-only}(3), we obtain an algorithm for relation-dominated queries.

\begin{theorem}
    Algorithm \ref{alg:R1} computes any relation-dominated query in $O(N)$ time.
\end{theorem}

For other queries, we proceed to the second round.

\subsection{Second-round computation}  
The second-round computation relies on the notion of \emph{dangling-free} relations and \emph{reducible} relations.

\begin{definition}[Dangling-free Relations]
Given a conjunctive query $\Q := \pi_{\O} \left(\Join_{k \in [n]} R_k\right)$, a relation $R_i$ is \emph{dangling-free} if on every database instance, for every $t \in R_i$, there exists a full join result $t' \in \Join_{k \in [n]} R_k$ such that $t = \pi_{\A_i} t'$.
\end{definition}

\begin{lemma}
\label{lem:free-root}
    For any acyclic query $\Q$ and any join tree $\T$ of $\Q$, the root node $R_r$ of $\T$ after the first round is dangling-free.
\end{lemma}

\begin{definition}[Reducible Relations]
    Let $\Q$ be an acyclic CQ and $\T$ be a join tree of $\Q$.  Consider a relation $R_i(\A_i) \in \T$, and let $R_j(\A_j)$ be a neighbor of $R_i$.
    We say that $R_j$ is \textit{reducible} for $R_i$ if, for every other neighbor $R_k(\A_k)$ of $R_i$, $\A_k \cap \A_i \subseteq \O$.
\end{definition}

As example, in the join tree of Figure \ref{fig:ex2_1_jt_2}, $R_1$ has only one reducible relation $R_3$; $R_4$ are $R_5$ are not reducible for $R_1$, because $R_3$ and $R_1$ have a non-output join attribute $x_4$.  

The following two special cases will be useful later: (1) For any leaf node, its parent is always reducible for it because it has no other neighbors. (2) In a full query, each node is reducible for all of its neighbors.

The second round revolves around dangling-free relations and their reducibles, since their joins have bounded size: 

\begin{lemma}
\label{lem:danglingbound}
    Given an acyclic query $\Q$ and a join tree $\T$ for $\Q$ after the first round, let $R_i$ be a dangling-free relation and $R_j$ be a reducible relation for $R_i$.  Then $|R_i \Join R_j| = 
    O(\min(N\OUT, F))$.  Furthermore, if $\bar{\A}_j \cap \A_j \subseteq \O$, then $|R_i \Join R_j|=O(\OUT)$.
\end{lemma}

In the second round, we iteratively identify any dangling-free relation $R_i$ and one of its reducible relations $R_j$, perform a join followed by a projection, and reduce the join tree by one relation, as shown in Algorithm~\ref{alg:reduction}.  In the algorithm, $\Delta$ represents symmetric  difference, i.e., $\A_i \Delta \A_j = (\A_i - \A_j) \cup (\A_j - \A_i)$. 

\begin{algorithm}[h]
    \caption{Reduction($\Q, \T, R_i, R_j$)}
    \label{alg:reduction}
    \KwIn{A join tree $\T$ for the acyclic query $\Q$ with relation $\bm{R}$, where $R_j$ is a reducible relation of a dangling-free relation $R_i$}
    \KwOut{A resulting join tree $\T'$ and a reduced query $\Q'$ on relations $\bm{R'}$, where $|\bm{R'}| = |\bm{R}| - 1$}
    
    $\T' \gets \T$, $\bm{R}' \gets \bm{R}$\;
    $R_i' \gets \pi_{\O \cup \left(\A_i \Delta \A_j\right)} \left(R_i \Join R_j\right)$\;
    $\bm{R}' \gets \left(\bm{R}' - \{R_i\} - \{R_j\}\right) \cup \{R_i'\}$\;
    In $\T$, merge $R_i$ and $R_j$ into $R_i'$\;
    $\Q' := \pi_{\O} \left(\Join_{R \in \bm{R'}} R\right)$\;
    \KwRet{$\Q', \T', \bm{R'}$}
\end{algorithm}

We need to show that a pair of dangling-free and reducible relations always exist, so that we can repeatedly apply Algorithm \ref{alg:reduction}.  It is easy to show that dangling-free relations always exist.  In particular, the root of the join tree after the first round must be dangling-free (Lemma \ref{lem:free-root}).  Also, the join of a dangling-free relation with any other relation must still be dangling-free, so the newly generated relation $R_i'$ by Algorithm \ref{alg:reduction} is also dangling-free.


However, it is not clear if reducible relations always exist.  We consider the following two cases separately.

\paragraph{Free-connex queries}
If $\Q$ is free-connex, then the query after the first round is full.  By the observation earlier, every relation is reducible to all its neighbors.  Thus we can apply Algorithm \ref{alg:reduction} on the root $R_r$ and any of its child $R_j$.  The newly generated relation is still dangling-free and it becomes the new root.  We can thus repeatedly apply Algorithm \ref{alg:reduction} until only one relation remains.  

In terms of running time, observe that in a full query, the second part of Lemma \ref{lem:danglingbound} applies, so the cost of each join is $O(N+M)$.  Combining with Lemma \ref{lem:firstround}, we conclude:


\begin{theorem}
    Algorithm \ref{alg:R1} and \ref{alg:reduction} compute any free-connex query in $O(N+\OUT)$ time.
\end{theorem}

\begin{example}
We continue Example \ref{ex:fc}. After the first-round computation, the join tree is shown in Figure \ref{fig:jointree2}, which is a full query.  The root $R_1$ is dangling-free, and both of its children $R_2$ and $R_4$ are reducible.  Applying Algorithm \ref{alg:reduction} twice yields the following query plan (continuing the plan in Example \ref{ex:fc}):
\[
\begin{aligned}
    (7) &\quad R_1 \gets R_1 \Join R_2; & (8) &\quad \Q \gets R_1 \Join R_4.  \qed
\end{aligned}
\] 
\end{example}





\paragraph{Non-free-connex queries}
Although dangling-free relations must exist for non-free-connex queries after the first round (at least, the root of $\T$ is one), but they may not have any reducible neighbors.  In this case, we use semi-joins to make additional relations dangling-free, based on the following lemma:

\begin{lemma}
\label{lem:childdangling}
    For any acyclic query $\Q$, let $\T$ be the join tree after the first-round computation.  Let $R_i$ be any dangling-free relation in $\T$, and $R_j$ be any child of $R_i$.  
    If we replace $R_j$ with $R_j' := R_j \ltimes R_i$, then the query is equivalent while $R_j'$ is dangling-free for $\Q$.
\end{lemma}

As observed earlier, when a leaf becomes dangling-free, its parent must be reducible, so this strategy can always succeed in finding a pair of relations to apply Algorithm \ref{alg:reduction}.   

\begin{example}
\label{ex:Q3Con}
    We continue Example~\ref{ex:Q3}.  The join tree is shown in Figure \ref{fig:jointree1} after the first round. The root $R_5$ is dangling-free, but neither of its children is reducible.
    Then we can use a semi-join to make $R_6$ dangling-free, and then apply Algorithm \ref{alg:reduction} to merge $R_5$ and $R_6$.  After this, $R_1$ becomes the only neighbor of $R_5$, hence reducible.  The query plan is (continuing the plan in Example \ref{ex:Q3}):
\[
\begin{aligned}
    (7) & \quad R_6 \gets R_6 \ltimes R_5; & (8) & \quad R_5 \gets \pi_{x_4, x_8}\left(R_5 \Join R_6\right); \\ 
    (9) &\quad \Q_1 \gets \pi_{x_1, x_2, x_8} \left(R_5 \Join R_1\right).  & &
\end{aligned}
\]

Compared with the original Yannakakis plan (Example \ref{ex:yan}), we see that our plan uses only 3 semi-joins as opposed to 10, and 3 aggregation-join operations have been pushed to before the semi-joins.  We ran the three plans in DuckDB on the 5-copy SF=100 TPC-H dataset, DuckDB's plan took 488 s, the original Yannakakis plan took 21.1 s, while our new plan took 13.2 s.
\qed
\end{example} 

Finally, we can show that our algorithm achieves the same running time guarantee as that of the Yannakakis algorithm for acyclic but non-free-connex queries:
\begin{theorem}
\label{thm:non-free-connex}
    Algorithm \ref{alg:R1} and \ref{alg:reduction} compute any acyclic query in $O(\min(N \OUT, F))$ time.
\end{theorem}

\section{General Queries}
\label{sec:general}
\subsection{Cyclic Queries}
\label{sec:cyclic}
Our previous discussions were based on acyclic CQs with a join tree.  For cyclic CQs, \textbf{Generalized Hypertree Decomposition (GHD)} \cite{Abo2017Shannon, gottlob99Hypertree} is a powerful tool for efficiently transforming them into acyclic CQs.  A GHD also takes the form of a tree $\T$, whose nodes are often called \textit{bags}. But unlike the join tree for acyclic queries that maps each node to a single relation, each node $\textsf{Bag}_j$ of $\T$ maps to a set of attributes $\mathcal{B}_j$, where (1) for every relation $R_i(\A_i)$, there exists a node $\textsf{Bag}_j$ such that $\A_i \subseteq \mathcal{B}_j$ and (2) for each attribute $x$, all nodes of $\T$ containing $x$ form a connected subtree of $\T$.  Such a tree $\T$ is called a \emph{generalized join tree}, and we said the tree is \emph{generalized free-connex join tree} if it also satisfies the free-connex condition.  Each bag $[\mathcal{B}]$ can be materialized by the following query:
\begin{equation}
\label{eq:bag}
    \Q_\mathcal{B} \gets \Join_{R(\A) \in \bm{R}, \A \subseteq \mathcal{B} \neq \emptyset}\left(R\right). 
\end{equation}
It should be noted that each relation can appear in multiple bags. In order to prevent miscalculations of the aggregate value, we create a special relation $R_i^\textbf{1}$ for each $R_i \in \mathcal{R}$. For each $t \in R_i$, we add $t$ to $R_i^\textbf{1}$ with the annotation $v(t) = \textbf{1}$. Then, we replace $R_i$ with $R_i^\textbf{1}$ for all bags except for one with $\A_i \subseteq \mathcal{B}$.  

In order to evaluate a CQ $\Q$ on the given generalized join tree $\T$, we start by materializing each bag $[\mathcal{B}]$. This involves evaluating $\Q_\mathcal{B}$ directly in the database with a binary join plan (or WCOJ if available) and then replacing the bag with the materialized relation $R_\mathcal{B}$. Our cost-based optimization further improves the pre-processing by selecting the best join orders for the binary join plan. Once this process is complete, the resulting tree becomes a normal join tree and can be evaluated directly using \name.

In this work, we adopt a similar approach to the previous state-of-the-art \cite{Aberger2017Empty}, which exhaustively explores all possible generalized hypertree decompositions (GHDs). Our cost-based optimizer enhances the efficiency of GHD search by employing our cost estimator to obtain more accurate results than the standard search algorithms that rely on heuristics. Additionally, when calculating the size of each GHD bag, we take cardinality constraints into account. For example, if a bag contains relations $R_1(x_1, x_2)$ and $R_2(\underline{x_2}, x_3)$, where $x_2$ is a primary key for $R_2$, we conceptually merge them into a new relation $R_{12}(x_1, x_2, x_3)$ with $|R_{12}| = |R_1|$. This approach provides a more accurate cost estimation, allowing our optimizer to select the most efficient GHD. 

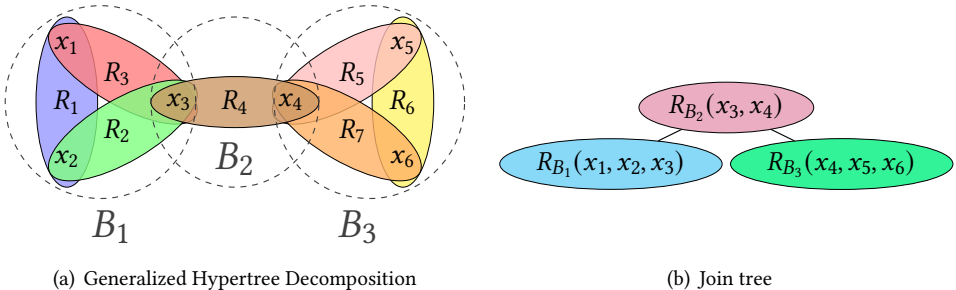
\begin{figure}[ht]
    \subfigure[Generalized Hypertree Decomposition]
    {
    \centering
    \begin{minipage}[t]{0.45\columnwidth}
        \centering
        \resizebox{\linewidth}{!}{
            \begin{tikzpicture}[font = \LARGE]
            \node (w1) at (-3, 0) {};
            \node (w2) at (-2, 0.5) {};
            \node (w3) at (-2, -0.5) {};
            \node (w4) at (3, 0) {};
            \node (w5) at (2, 0.5) {};
            \node (w6) at (2, -0.5) {};
            
            \begin{scope}[fill opacity=0.8]
            \filldraw[fill=yellow!70, rotate=90] ($(w4)$) ellipse (1.5 and 0.55);
            \filldraw[fill=pink, rotate=30] ($(w5)$) ellipse (1.5 and 0.55);
            \filldraw[fill=orange!60, rotate=150] ($(w6)$) ellipse (1.5 and 0.55);
            \filldraw[fill=blue!40, rotate=90] ($(w1)$) ellipse (1.5 and 0.55);
            \filldraw[fill=red!50, rotate=150] ($(w2)$) ellipse (1.5 and 0.55);
            \filldraw[fill=green!50, rotate=30] ($(w3)$) ellipse (1.5 and 0.55);
            \filldraw[fill=brown!80] ($(0, 0)$) ellipse (1.5 and 0.45);
            \end{scope}

            \node (v1) at (-3,1) {$x_1$};
            \node (v2) at (-3,-1) {$x_2$};
            \node (v3) at (-1,0) {$x_3$};
            \node (v4) at (1,0) {$x_4$};
            \node (v5) at (3,1) {$x_5$};
            \node (v6) at (3,-1) {$x_6$};

            \node at (-3, 0) {$R_1$};
            \node at (-2.1, 0.5) {$R_3$};
            \node at (-2.1,-0.5) {$R_2$};
            \node at (3, 0) {$R_6$};
            \node at (2.1,0.5) {$R_5$};
            \node at (2.1,-0.5) {$R_7$};
            \node at (0, 0) {$R_4$};

            \draw[dashed, line width=0.2mm, darkgray] (-2.4,0) circle (1.7);
            \node at (-2.2,-2.2) {\textbf{\textcolor{darkgray}{\Huge $B_1$}}}; 

            \draw[dashed, line width=0.2mm, darkgray] (0,0) circle (1.5);
            \node at (0,-1) {\textbf{\textcolor{darkgray}{\Huge $B_2$}}}; 

            \draw[dashed, line width=0.2mm, darkgray] (2.4,0) circle (1.7);
            \node at (2.2,-2.2) {\textbf{\textcolor{darkgray}{\Huge $B_3$}}}; 

            \end{tikzpicture}
        }
        \label{fig:ghd1}
    \vspace*{-0.2cm}
    \end{minipage}
    }
    \subfigure[Join tree]
    {
    \begin{minipage}[t]{0.44\columnwidth}
        \centering
        \raisebox{0.8cm}{
            \resizebox{\linewidth}{!}{
                \begin{tikzpicture}[font=\Huge, every node/.style={draw, ellipse, minimum width=12mm, minimum height=8mm, ultra thin}]
                    \node[fill=cyan!50, fill opacity=0.8, text opacity=1] (B1) at (-3.1,-1.5) {$R_{B_1}(x_1, x_2, x_3)$};
                    \node[fill=purple!40, fill opacity=0.8, text opacity=1] (B2) at (0,0) {$R_{B_2}(x_3, x_4)$};
                    \node[fill={rgb,255:red,0; green,240; blue,127}, fill opacity=0.8, text opacity=1] (B3) at (3.1,-1.5) {$R_{B_3}(x_4, x_5, x_6)$};
                    \draw[line width=0.3mm] (B2) -- (B3);
                    \draw[line width=0.3mm] (B2) -- (B1);
                \end{tikzpicture}
            }
        }
        \label{fig:ghd2}
    \vspace*{-0.2cm}
    \end{minipage}
    }
    \vspace*{-0.3cm}
    \caption{An Example of GHD and its acyclic CQ.}
    \label{fig:ghd}
\end{figure}
\begin{example}

See Figure~\ref{fig:ghd1} as an example of GHD on a natural join of 7 relations:
$R_1(x_1, x_2)$, $R_2(x_2, x_3)$, $R_3(x_3, x_1)$, $R_4(x_3, x_4)$, $R_5(x_4, x_5)$, $R_6(x_5, x_6)$ and $R_7(x_6, x_4)$.
There are three bags in the decomposition:
$$R_{\mathcal{B}_1} \gets R_1(x_1, x_2) \Join R_2(x_2, x_3) \Join R_3(x_3, x_1);$$
$$R_{\mathcal{B}_2} \gets R_4(x_3, x_4);$$
$$R_{\mathcal{B}_3} \gets R_5(x_4, x_5) \Join R_6(x_5, x_6) \Join R_7(x_6, x_4).$$
After performing two triangle joins ($R_{\mathcal{B}_1}$ and $R_{\mathcal{B}_3}$) we get an acyclic join (line-3 join) as Figure~\ref{fig:ghd2}.  The tree can be evaluated in a total time of $O(N^{1.5}+\OUT)$ with worst-case optimal joins, or $O(N^{2}+\OUT)$ in most industrial database systems.
    
\end{example}

\subsection{Sub-queries, Unions, Differences, and Top-k} The support for other operations in DBMS on our newly developed algorithm is natural by considering the underlying conjunctive query as a special relation.  The evaluation and materialization of the underlying conjunctive query can be done by using the new algorithm.  Then, additional operators can be applied to the query results by replacing the conjunctive queries with the new relation.   
\begin{example}
Consider the TPC-H Benchmark Query 17:
\begin{lstlisting}
SELECT SUM(l_extendedprice) / 7.0 as avg_yearly
FROM lineitem, part
WHERE p_partkey = l_partkey and p_brand = 'Brand#23' and p_container = 'MED BOX' and l_quantity < (
    SELECT 0.2 * avg(l_quantity) FROM lineitem WHERE l_partkey = p_partkey);
\end{lstlisting}
which is a nested query.  To evaluate the nested query, our framework will first evaluate the underlying conjunctive query
\begin{lstlisting}
SELECT 0.2* avg(l_quantity) as cnt FROM lineitem, part
WHERE p_partkey = l_partkey and p_brand = 'Brand#23' and p_container = 'MED BOX';
\end{lstlisting}
then using the query result $R_Q$ as a new input relation, and evaluate another conjunctive query
\begin{lstlisting}[mathescape=true]
SELECT SUM(l_extendedprice) / 7.0 as avg_yearly
FROM lineitem, part, $R_Q$
WHERE p_partkey = l_partkey and p_brand = 'Brand#23' and p_container = 'MED BOX' and l_quantity < cnt;
\end{lstlisting}
\end{example}
While this allows our algorithm to support universal SQL queries and our algorithm can guarantee an output-sensitive running time when evaluating the conjunctive queries, we cannot guarantee an output-sensitive running time for the entire query as the output size of these conjunctive queries can be significantly larger than the final size of the query results.  Recent advances have shown that we can push down unions\cite{Nofar2021Union}, differences\cite{Hu2023Computing}, and Top-k\cite{WANG2024Relational} while evaluating those conjunctive queries. As a natural aspect of relational algorithms, our framework can be extended to support all these queries with additional rewrite steps and only add a constant or logarithmic cost to the complexity.

\begin{example}
    Consider the following difference of conjunctive query (DCQ) studied in \cite{Hu2023Computing}:
    \[
        \pi_{x_4} \left(R_1(x_1, x_2) \Join R_2(x_2, x_3, x_4) - R_3(x_1, x_2, x_3) \Join R_4(x_3, x_4)\right),
    \]
One way to evaluate the query is by first evaluating the two queries, $R_1 \Join R_2$ and $R_3 \Join R_4$, and then calculating the difference between the two queries and performing the projection. However, it is possible that the final result is empty, even though the two conjunctive queries can produce $O(N^2)$ results in the worst case. By using the techniques introduced in \cite{Hu2023Computing}, we can rewrite the process as follows:
    \[
        \begin{aligned}
         & \pi_{x_4} \left(R_1 \Join R_2- R_3 \Join R_4\right)\\
         = & \pi_{x_4} \left(\left(R_1 - \pi_{x_1, x_2} R_3\right) \Join R_2\right) \cup \pi_{x_4} \left(R_1 \Join \left(R_2 - \left(\pi_{x_2, x_3} R_3\right) \Join R_4\right)\right),   
        \end{aligned}
    \]
    and 
    \[
        \begin{aligned}
        & \left(R_2 - \left(\pi_{x_2, x_3} R_3\right) \Join R_4\right) \\
            = & R_2 \ltimes \left(\pi_{x_2, x_3} R_2 - \pi_{x_2, x_3} R_3\right) \cup R_2 \ltimes \left(\pi_{x_3, x_4} R_2 - R_4\right)
        \end{aligned}
    \]
    where each individual query's output size is bounded by the actual output size of the DCQ. With \name, those queries can be evaluated in $O(N+\OUT)$ time, where $\OUT$ represents the actual output of the DCQ.
\end{example}

\section{Query Optimization}
\label{sec:opt}
\name provides the same asymptotic running time guarantee with any valid join tree (free-connex join tree, or relation-dominated join tree, respectively).  However, there are still constant-factor differences between these join trees; even for the same join tree, different reduction orders during the two rounds of computations can also make some differences. 

\begin{example}
\label{ex:jt}
Consider $\Q_1$ from Example~\ref{ex:query}, where the new query plan, using the join tree $\T_1$, significantly improves the performance compared with the original query plan in DuckDB. However, our query optimizer can find a better join tree $\T_3$ by simply rotating the tree with $R_1$ as the root node, as shown in Figure \ref{fig:jointree3}. Despite the small changes to the join tree, the new query plan reduces the total intermediate results from 556,473,531 to 242,661,000 and eliminates one semi-join step in the second round of computation. The resulting running time on $\T_3$ is 6.800 s, which is approximately 49\% less compared to the plan on $\T_1$. \qed

\begin{figure}[h]
    \centering
    \subfigure{
        \begin{minipage}[t]{0.48\columnwidth}
        \centering
        \resizebox{\linewidth}{!}{
            \begin{tikzpicture}[
            level 1/.style={sibling distance=24mm, level distance=11mm},
            level 2/.style={sibling distance=24mm, level distance=11mm},
            every node/.style={draw, ellipse, minimum width=10mm, minimum height=6mm, font=\normalsize, ultra thin}]
            \node[fill={rgb,255:red,135; green,206; blue,250}, fill opacity=1, text opacity=1] (R1) at (0, 0) {$R_1(\underline{x_1}, \underline{x_2}, x_3, x_4)$}
                child {node[fill={rgb,255:red,255; green,138; blue,128}, fill opacity=1, text opacity=1] (R3) {$R_3(x_3, x_4)$}
                    child {node[fill={rgb,255:red,225; green,190; blue,231}, fill opacity=1, text opacity=1] (R4) {$R_4(x_3, x_6)$}}
                }
                child {node[fill={rgb,255:red,255; green,205; blue,210}, fill opacity=1, text opacity=1] (R5) {$R_5(x_4, x_7)$}
                    child {node[fill={rgb,255:red,255; green,255; blue,102}, fill opacity=1, text opacity=1] (R6) {$R_6(x_7, \underline{x_8})$}}
                }
                child {node[fill={rgb,255:red,80; green,255; blue,80}, fill opacity=1, text opacity=1] (R2) {$R_2(\underline{x_2}, x_5)$}};
        \end{tikzpicture}
        }
        \caption{Join tree $\T_3$ for $\Q_1$.}
        \label{fig:jointree3}
        \end{minipage}
    }
    \subfigure{
        \begin{minipage}[t]{0.4\linewidth}
        \centering
        \resizebox{\linewidth}{!}{
            \begin{tikzpicture}[
    every node/.style={draw, ellipse, minimum width=10mm, minimum height=6mm, font=\normalsize, ultra thin},
    node distance=22mm,
    ultra thin
]

    \node (customer) [fill={rgb,255:red,225; green,190; blue,231}] at (1.5, -2) {$R_3(\overline{x_3}, x_4')$};
    \node (orders) [fill={rgb,255:red,255; green,138; blue,128}] at (1.5, -1) {$R_2(\overline{x_2}, x_3, x_8)$};
    \node (lineitem) [fill={rgb,255:red,135; green,206; blue,250}] at (0, 0) {$R_1(x_1, x_2)$};
    \node (supplier) [fill={rgb,255:red,80; green,255; blue,80}] at (-1.5, -1) {$R_5(\overline{x_1}, \underline{x_4})$};
    \node (nation) [fill={rgb,255:red,255; green,255; blue,102}] at (-1.5, -2) {$R_4(\overline{\underline{x_4}}, x_5, x_6)$};
    \node (region) [fill={rgb,255:red,255; green,205; blue,210}] at (-1.5, -3) {$R_6(\overline{x_6}, x_{7})$};

    \draw[-] (lineitem) -- (supplier);
    \draw[-] (supplier) -- (nation);
    \draw[-] (nation) -- (region);
    \draw[-] (lineitem) -- (orders);
    \draw[-] (orders) -- (customer);

    \end{tikzpicture}
    }
    \caption{Join tree for $\Q_5'$.}
    \label{fig:q5}
    \end{minipage}}
    \label{fig:opt}
\end{figure}

        
\setcounter{figure}{7} 

\end{example}

Thus, it is practically important to choose an optimal (or near-optimal) query plan from this family of plans.  We have designed a query optimizer tailored for \name, which consists of a rule-based component and a cost-based component, described below.

\subsection{Rule-Based Optimization}

\paragraph{Cycle Elimination. }  \name is designed to process acyclic queries and use GHD to transform cyclic queries into acyclic with extra cost.  However, some queries, although cyclic, can be turned into acyclic without affecting the running time by exploiting the PK constraints.  
\begin{example}
TPC-H query 5 can be represented as the following conjunctive query
\[
    \begin{aligned}
    \Q_5 \gets & \pi_{x_{5}} R_1(x_1, x_2) \Join R_2(\bar{x}_2, x_3, x_8) \Join R_3(\bar{x}_3, x_4) \\ 
    &\Join R_4(\bar{x}_4, x_5, x_6) \Join R_5(\bar{x}_1, x_4) \Join  R_6(\bar{x}_6, x_7),
    \end{aligned}
\]
where all primary keys are marked as $\bar{x}$.  The query is not acyclic due to the cycle created by $R_1, R_2, R_3, R_5$. We break the cycle by renaming one of the $x_4$'s into $x_4'$, but then reinforcing it by a selection:
\[
    \begin{aligned}
    \Q_5' \gets & \sigma_{x_4=x_4'}\Big(\pi_{x_{5}, x_{4}, x_4'} R_1(x_1, x_2)  \Join R_2(\bar{x}_2, x_3, x_8) \Join R_3(\bar{x}_3, x_4')\\
    &\Join R_4(\bar{x}_4, x_5, x_6) \Join  R_5(\bar{x}_1, x_4) 
    \Join R_6(\bar{x}_6, x_7)  \Big).
    \end{aligned}
\]
Now the query (before the selection $\sigma_{x_4=x_4'}$) is acyclic, with a join tree shown in Figure \ref{fig:q5}.  Meanwhile, since all joins are PK-FK joins, all intermediate join sizes are bounded by $O(N)$, so the overall running time is still $O(N)$, including the last selection step $\sigma_{x_4=x_4'}$.  \qed
\end{example}

\paragraph{Aggregation Elimination. } The PK constraint (in fact, any UNIQUE constraint) can help remove some redundant aggregations.  When the group-by attribute is a PK, the aggregation can be eliminated, i.e., line 5 and 9 in 
Algorithm~\ref{alg:R1}. 

\paragraph{Semi-join Elimination. } For a leaf relation $R$ and its parent node $R_p$ on the join tree, if the join key is a primary key of $R$ and foreign key of $R_p$, and there is no filtering condition on $R$, then the semi-join step between $R$ and $R_p$ can be ignored. This is because the PK-FK relationship already ensures that all tuples can be joined.

\begin{example}
Consider the query $\Q_1$ from Example~\ref{ex:Q3}. In the first round, $R_2$ was projected to $x_2$ before joining with $R_1$ to avoid duplication. However, if $x_2$ is a primary key of $R_2$, the projection is unnecessary.  In addition, if the PK-FK relationship holds between $R_5$ and $R_6$, the semi-join in Step (1) and Step (7) can be omitted without increasing the complexity.
\end{example}


\paragraph{Pruning for Annotation. }  In Section~\ref{sec:prelimiary}, the definition of conjunctive queries requires an additional annotation column for each relation to support the calculation of aggregation functions. This helps to generalize the definition to accommodate various aggregations. However, in some cases, this annotation may be redundant, but our optimizer is designed to identify these cases and avoid the additional cost. Our experiments demonstrate the importance of this optimization for database systems with column-store.

\begin{example}
Consider the query from Example~\ref{ex:query}. If we change the corresponding aggregation function from SUM to MAX on $\text{ps\_supplycost} * \text{l\_quantity}$ to obtain the maximum cost, the query will be defined over the semiring $(\mathbb{R}, \max, \cdot)$. In this case, we won't need to assign additional annotations on relations except for Partsupp and Lineitem. Our optimizer detects such situations and eliminates those annotations from our plan.
\end{example}

\paragraph{Fusion of Dimension Relations.} When a query involves joins between a large relation and multiple small relations, the optimizer can enhance efficiency by first join the small relations, or even using Cartesian products if they lack common attributes. This is because join or semi-join with the large relation can be more costly than performing a Cartesian product of the small relations.  For example, in the query $ R_1(x_1) \Join R_2(x_1, x_2) \Join R_3(x_2) $, if $|R_1|$ and $|R_3|$ are significantly smaller than $|R_2|$, we first perform the Cartesian product  $R_1 \times R_3$. Then, we apply our new query plan, which saves one join or semi-join with the large relation $R_2$.

\subsection{Cost-Based Optimization}
\label{sec:CBO}

Cost-based optimization in database systems is a key technique for enhancing query performance and resource usage. Our new algorithms specifically focus on the efficiency of a query plan within the algebraic structure. For all valid join trees, they have the same theoretical worst-case complexity. Therefore, it's important for us to take into account instance-specific information in order to identify the best query plan among all available options. In contrast to the standard binary join approach, which may not perform well due to the amplification of errors by join operations, operators like semi-join have a bounded cost that does not amplify errors. Additionally, the linear time guarantee provides an upper bound on the cost estimation. These factors make the standard cost-based optimization more effective for our new query plans.

\paragraph{Plan Enumeration (PE)} 
The first step in plan enumeration is to generate all valid join trees for the given query. For acyclic queries, we use GYO reduction \cite{university1980universal, yu1979algorithm} to enumerate all valid join trees. However, for cyclic hypergraphs, directly applying GYO reduction cannot reduce the query to an empty graph. Therefore, we compute all possible generalized hypertree decompositions (GHDs) \cite{gottlob2009generalized}. 

After generating a set of valid plans, we employ the following pruning strategies to control their number:
\begin{itemize}[leftmargin=*] 
\item For queries with output attributes, we require the root node to contain output attributes; 
\item We prefer plans where the larger relations are at the top of the tree; 
\item Unlike current database optimizers that tend to favor left-deep plans, we prioritize bushy plans with lower heights. 
\end{itemize} 
These rules help avoid additional costs when propagating large relations through intermediate results and make it easier for child nodes to prune their parent nodes.

\paragraph{Cardinality Estimation (CE) and Cost Model (CM)} Estimating the cardinality of intermediate results in a query plan has been extensively studied in the literature.  \revise{Thanks to their theoretical guarantee, the \name plans are less sensitive to CE/CM than traditional plans.  Bad CE/CM leads to at most a constant-factor difference for \name, while they may incur a polynomial-factor degradation for traditional plans, from $O(N)$ to $O(N^2)$ or even worse.  We used the standard CE and CM methods to ensure the best database engine compatibility and fair comparison, while better CE/CM may further improve the performance of \name.}  We first collect basic statistical information from the base tables, including their size, the number of distinct values, the quantiles, etc.  Then, during query optimization, we estimate the join size, projection size, and selectivity of selection predicates using some classical methods \cite{job, joinsizeEst, sample-ndv,selectivity1, selectivity2}.  Finally, we convert the cardinality estimates into an estimate of the actual running time using a standard cost model.

\section{System implementation}
\label{sec:system}

\begin{figure}[htb]
    \centering
    \scalebox{0.65}{
    \includegraphics[width=\textwidth]{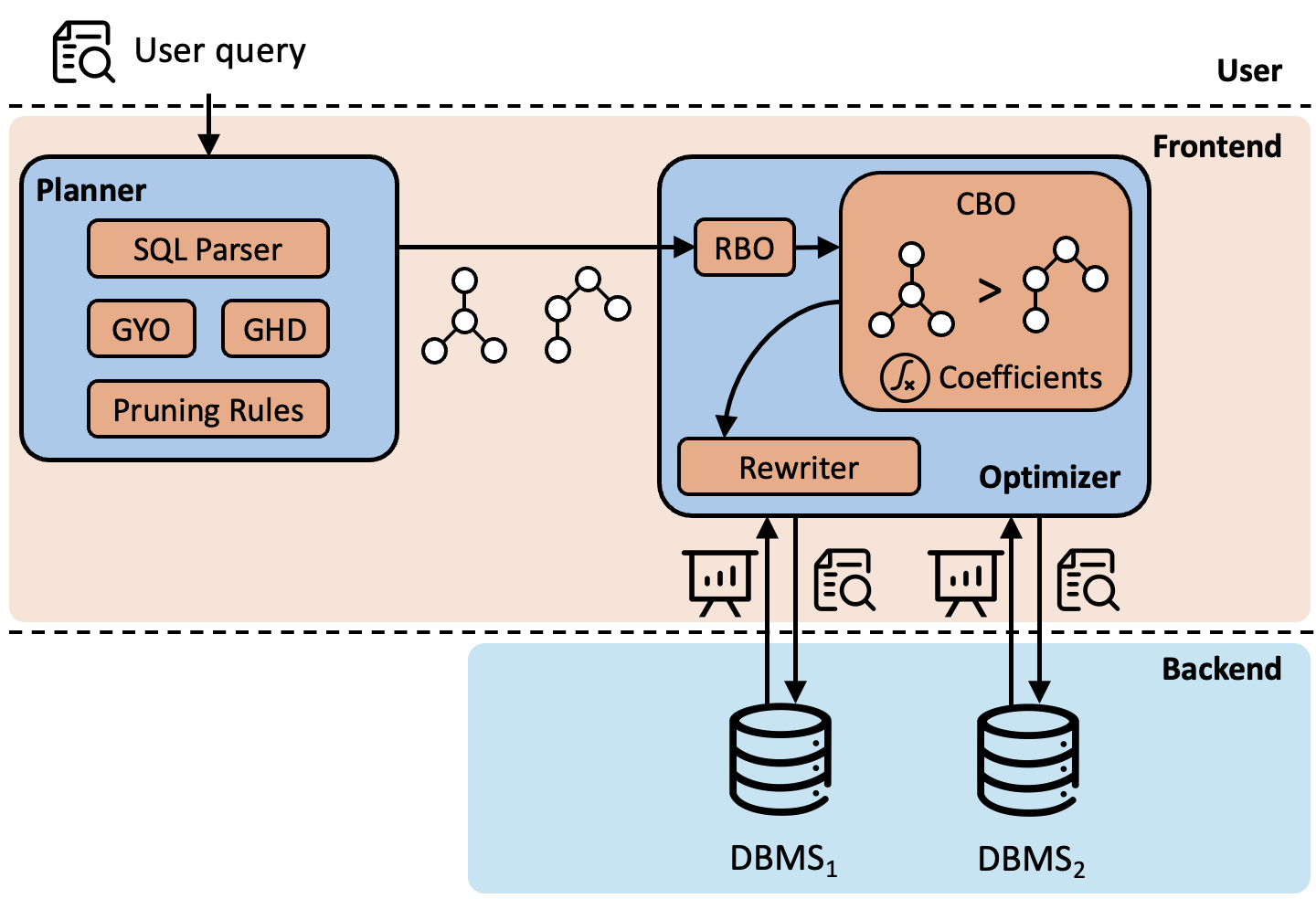}}
    \caption{System Architecture}
    \label{fig:system}
\end{figure}

We have developed a prototype system implementing our algorithms, which consists of two main components: the planner and the optimizer. Figure~\ref{fig:system} illustrates the architecture of our system. The planner accepts SQL queries and the database schema via REST APIs. Each input query first undergoes syntax validation using a built-in SQL parser based on Apache Calcite~\cite{calcite}. After validation, the query is transformed into a tree of relational operators. The planner then applies optimizations where appropriate, such as cycle elimination. Next, it builds the hypergraph and generates candidate join trees using the techniques described in Section~\ref{sec:CBO}. For free-connex queries, each candidate join tree is associated with a subtree $\T_n$, representing the connex subset. 

Upon receiving the candidate join trees, the optimizer uses a built-in cost model, along with statistics from the DBMS, to select the optimal join tree. In practice, the planning and optimization steps can be completed within 100 milliseconds. For complex queries that require longer optimization times, our system can choose to skip planning and optimization steps and directly use the join tree provided by the DBMS for the subsequent rewrite step, thereby balancing optimization time and query execution time.

For the rewrite step, our system employs the algorithm described in Section~\ref{sec:acyclic} to generate a series of equivalent intermediate representations (IRs) as instructions. Depending on the target DBMSs, the instructions are further converted into executable SQL queries. This design decouples our system from the underlying DBMSs, improving its portability. To support a new target DBMS, only the conversion from rewritten instructions to SQL statements is required. This also allows us to leverage some features of a specific DBMS for tailored performance optimization, enhancing runtime performance. For example, in DuckDB, we utilize temporary views to store the intermediate results of our plan, a method that allows it to follow our algorithm while introducing minimal overhead.  \revise{Each of the generated SQL statements in our plan is atomic and cannot be further optimized or rewritten.  We also verified in the experiments that all these systems executed the given plans as instructed.}

Our prototype is available at \cite{GitRepo}, currently supporting DuckDB \cite{duckdb}, PostgreSQL~\cite{postgre}, DBMS X (a commercial column-oriented database optimized for analytical processing), and SparkSQL~\cite{spark}.
\pdfimageresolution=1200


\section{Experimental Evaluation \& Analysis}
\label{sec:exp}

\subsection{Experimental Setup}

\noindent\textbf{Experimental Environment.}
Experiments for DuckDB and PostgreSQL were conducted on a machine with an Intel Xeon Gold 6354 CPU @ 3.00GHz (36 cores, 72 threads), 1TB RAM, running Ubuntu 20.04. The software versions used were DuckDB 1.0 and PostgreSQL 16.2. Spark experiments were performed on a machine with an Intel Xeon Silver 4116 CPU @ 2.10GHz (24 cores, 48 threads), 192GB RAM, running AlmaLinux 9.4, using Spark 3.5.1 with Java 1.8.0.

Each query was executed 10 times on each database engine, and we reported the median running time, \revise{including both optimization and execution time}. I/O time is excluded from the running time. Before running a query, we warm up the database and read all required relations into the memory.  A two-hour time limit was set for the SGPB and LSQB benchmarks, and a 30-minute limit for the TPC-H and JOB benchmarks. All systems were used with default configurations, utilizing all available resources: 72 threads for DuckDB and PostgreSQL, and 48 threads for SparkSQL.

\noindent\textbf{Datasets, Queries, and Benchmarks.} We assessed our algorithms using a variety of benchmarks covering graphs, social networks, and relational data to ensure a comprehensive evaluation across different data complexities and join types. All SQL queries used in our evaluation are available in our repository~\cite{GitRepo}.
\begin{itemize}[leftmargin=*] \item \textbf{Sub-Graph Pattern Benchmark (SGPB).} We designed queries over diverse graph datasets from the Stanford Network Analysis Project (SNAP)~\cite{SNAP}, including \textit{bitcoin}, \textit{epinions}, \textit{dblp}, \textit{google}, and \textit{wiki}, containing 24K to 28M edges. These datasets provide a robust test for graph query performance.

\item \textbf{LSQB.} The LSQB Benchmark~\cite{LSQB}, derived from the LDBC Social Network Benchmark (LDBC-SNB)~\cite{LDBC}, focuses on complex queries involving numerous joins typical in social network analysis. We evaluated all nine queries using a scale factor of 30.

\item \textbf{TPC-H.} TPC-H~\cite{tpch} is an industry-standard benchmark simulating decision support systems with large data volumes and complex queries addressing critical business questions. We conducted experiments using a scale factor of 100.

\item \textbf{JOB.} The Join Order Benchmark (JOB)~\cite{job} comprises 113 analytical queries over the Internet Movie Database (IMDB) dataset~\cite{imdb}. To illustrate performance improvements, we scaled the dataset by enlarging each table 10 to 100 times its original size.

\item \textbf{CEB.} The Cardinality Estimation Benchmark (CEB)~\cite{CEB, CEB2} is a benchmark consisting of millions of SQL queries, designed to test the performance of query optimization. It primarily features two workloads: IMDB and StackExchange. Similarly, we have scaled the dataset to 10 times its original size.
\end{itemize}

For all benchmarks, we focused on evaluating conjunctive queries with aggregations. We omitted operations like \texttt{LIMIT} or \texttt{ORDER BY} and replaced anti-joins or outer joins with inner joins to standardize the query patterns.

\subsection{Results}

\subsubsection{\textbf{Running Time Comparison}}

\begin{figure*}[htbp]
    \centering
    \begin{minipage}[t]{\textwidth}
        \centering
        \begin{minipage}[b]{\textwidth}
            \centering
            \includegraphics[width=\textwidth]{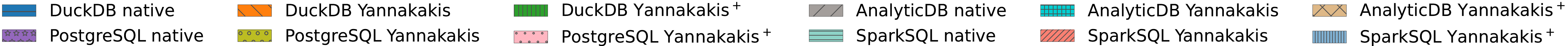}
            \label{fig:graph0}
            \vspace*{-0.45cm}
        \end{minipage}
        \begin{minipage}[b]{\textwidth}
            \centering
            \includegraphics[width=\textwidth]{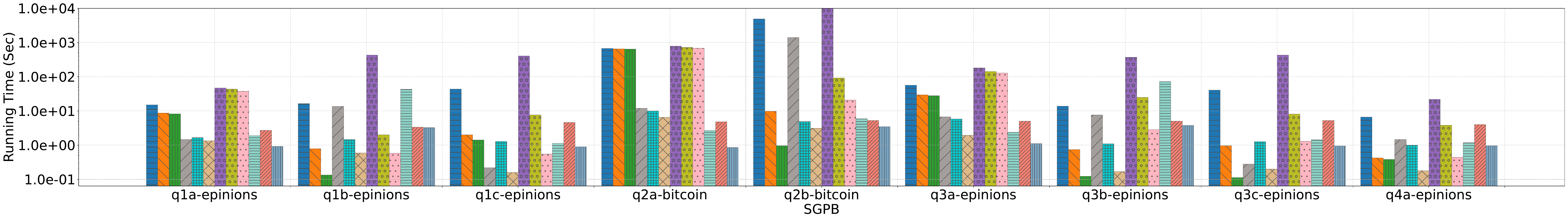}
            \label{fig:graph1}
            \vspace*{-0.45cm}
        \end{minipage}
        \begin{minipage}[b]{\textwidth}
            \centering
            \includegraphics[width=\textwidth]{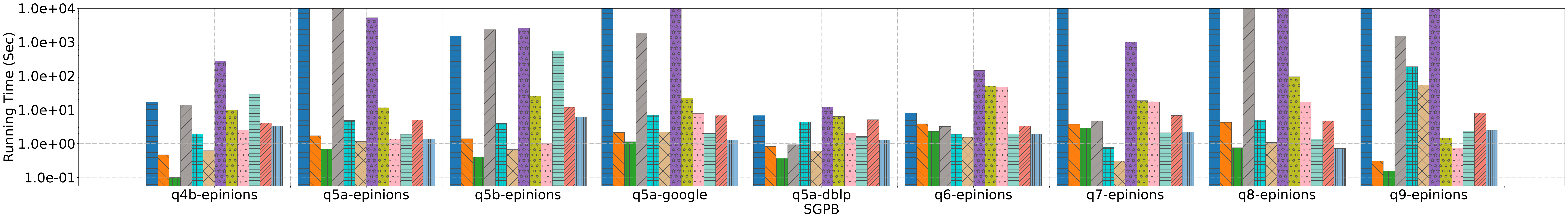}
            \label{fig:graph2}
            \vspace*{-0.45cm}
        \end{minipage}
        \begin{minipage}[b]{\textwidth}
            \centering
            \includegraphics[width=\textwidth]{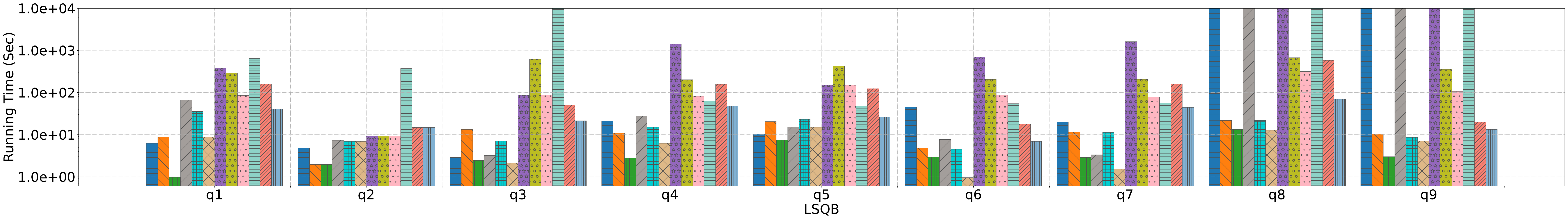}
            \label{fig:lsqb}
            \vspace*{-0.45cm}
        \end{minipage}
        \begin{minipage}[b]{\textwidth}
            \centering
            \includegraphics[width=\textwidth]{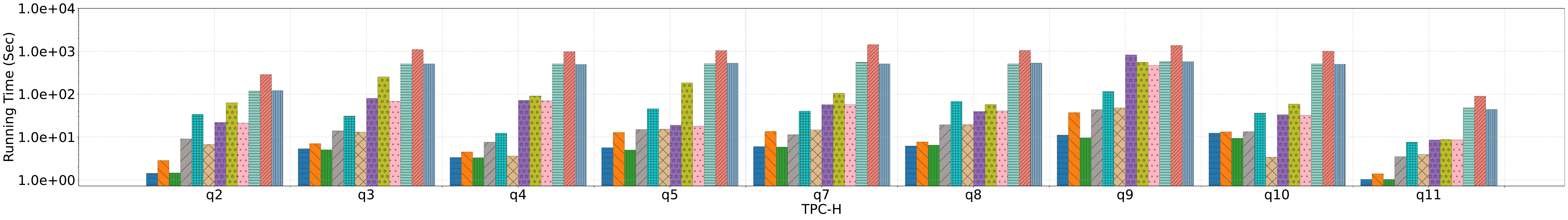}
            \label{fig:tpch1}
            \vspace*{-0.45cm}
        \end{minipage}
        \begin{minipage}[b]{\textwidth}
            \centering
            \includegraphics[width=\textwidth]{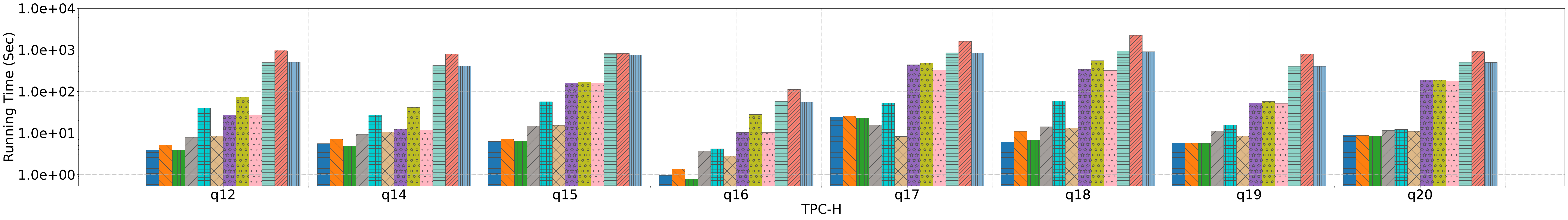}
            \label{fig:tpch2}
            \vspace*{-0.45cm}
        \end{minipage}
    \end{minipage}
    \vspace{1em}  
    \begin{minipage}[t]{0.5\textwidth}
        \hspace*{-0.7cm}  
        \includegraphics[width=\textwidth]{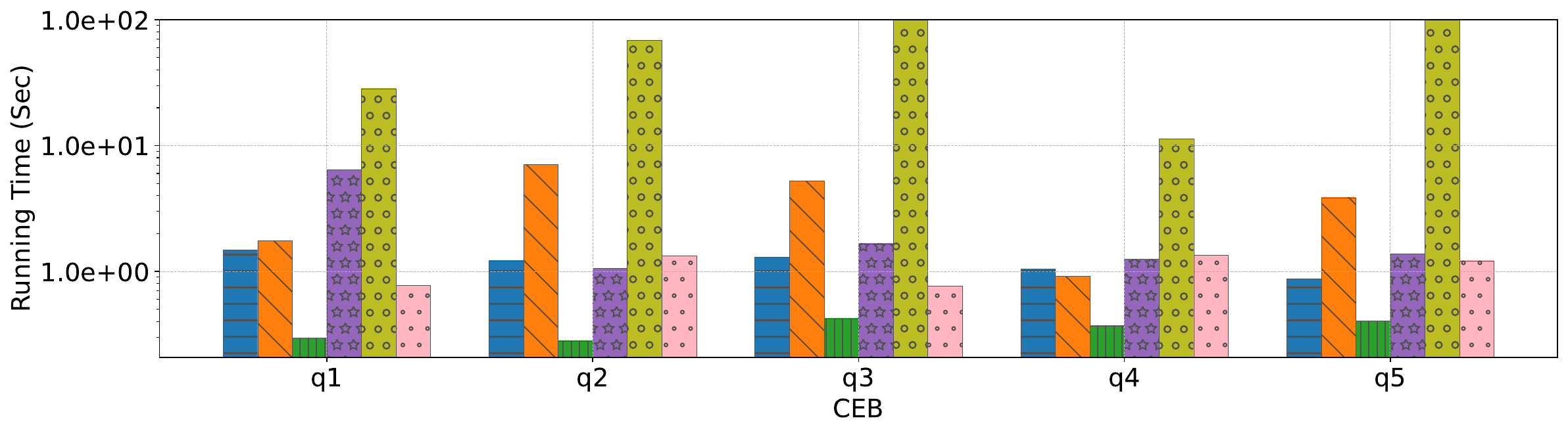}
        \vspace{-0.5cm}
    \end{minipage}
    \begin{minipage}[t]{0.4\textwidth}
        \raggedright  
        \vspace{-1.8cm}  
        \captionsetup{justification=centering,singlelinecheck=true}
        \caption{Running times of DuckDB, AnalyticDB, PostgreSQL, SparkSQL}
        \label{fig:all_benchmark}
    \end{minipage}
    \vspace{-0.5cm}
\end{figure*}

Figures~\ref{fig:all_benchmark} 
present the running times and relative speedups of our query rewriter across four benchmarks—SGPB, LSQB, TPC-H, and JOB—evaluated on DuckDB, AnalyticDB, PostgreSQL, and SparkSQL.  All bars reaching the axis boundary indicate that the system either exceeded the time limit or encountered memory issues.  All the raw experimental results available in our code repository~\cite{GitRepo}.

We first observe that, for the Yannakakis query plans, although they can significantly improve performance by orders of magnitude on queries like SGPB-q4b (35.83x) or SGPB-q5b (1071.78x), they yield significant performance drawbacks on plenty of queries. Especially for queries with PK-FK joins, 
when executing the Yannakakis plan on the JOB, most queries run slower than their native query plans. Such performance matches the previous observations \cite{ThomasPODS24, gottlob2023structure}, and the limited improvements are due to (1) The overhead introduced by splitting queries into multiple SQL statements and creating temporary views offsets potential gains. (2) Primary key–foreign key (PK-FK) constraints resulting in intermediate result sizes of $O(N)$, matching the time complexity of the Yannakakis algorithm and leaving little room for optimization.   

On the other hand, we observe significant performance improvement in our \name plan.  In the total \cbn{162} test queries across all platforms/benchmarks, we can achieve performance improvement over \cbn{160} queries 
compared with the native query plans, with an average of 2.4x and a maximum of 47,059x improvement. The performance drawbacks are limited, with $12.75\%$ additional running time at most on the test queries.  In addition, we achieved performance improvement over all queries compared with the Yannakakis query plans, with an average of \cbn{2.74x} and a maximum of \cbn{156.03x} improvement. The detailed results are:

\begin{figure}[t]
    \centering
    \renewcommand{\baselinestretch}{1.0}
    \hspace*{1cm}
     \begin{minipage}[b]{0.6\textwidth}
        \centering
        \includegraphics[width=\textwidth]{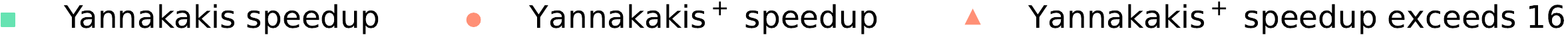}
        \vspace*{-0.4cm}
    \end{minipage}
    \begin{minipage}[b]{\textwidth}
        \centering
        \includegraphics[width=\textwidth]{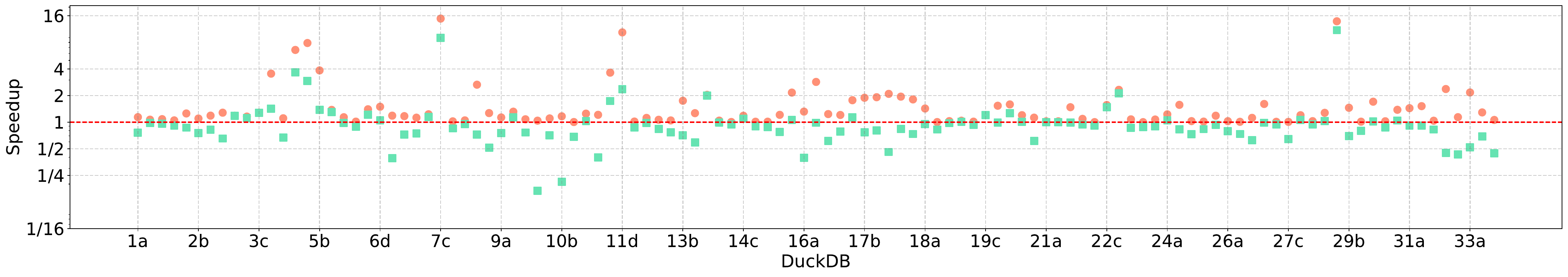}
        \label{fig:duckdb}
        \vspace*{-0.5cm}
    \end{minipage}
    \begin{minipage}[b]{\textwidth}
        \centering
        \includegraphics[width=\textwidth]{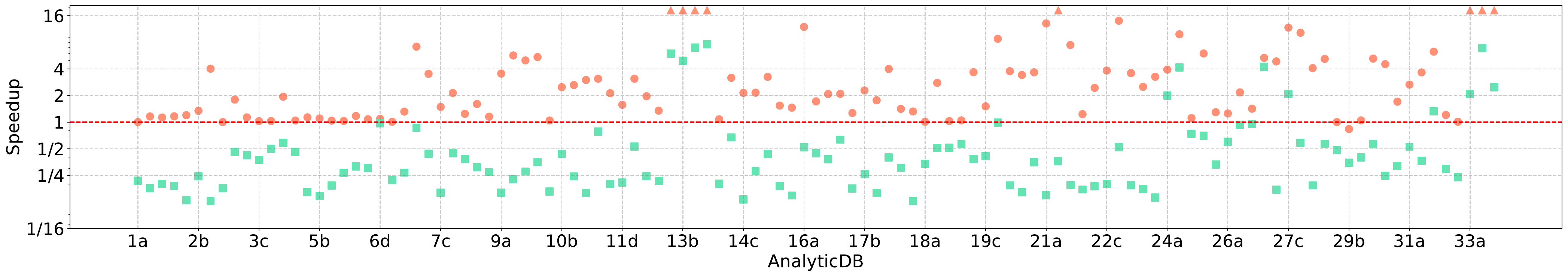}
        \label{fig:adb}
        \vspace*{-0.5cm}
    \end{minipage}
    \begin{minipage}[b]{\textwidth}
        \centering
        \includegraphics[width=\textwidth]{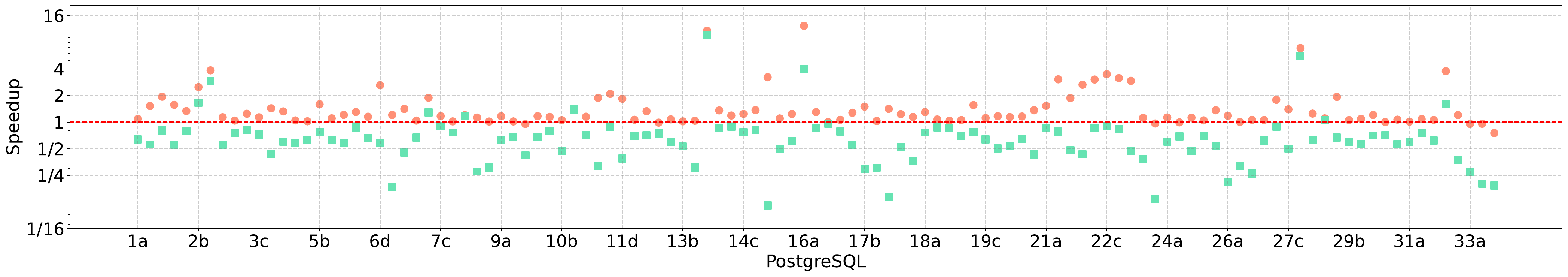}
        \label{fig:pg}
        \vspace*{-0.5cm}
    \end{minipage}
    \begin{minipage}[b]{\textwidth}
        \centering
        \includegraphics[width=\textwidth]{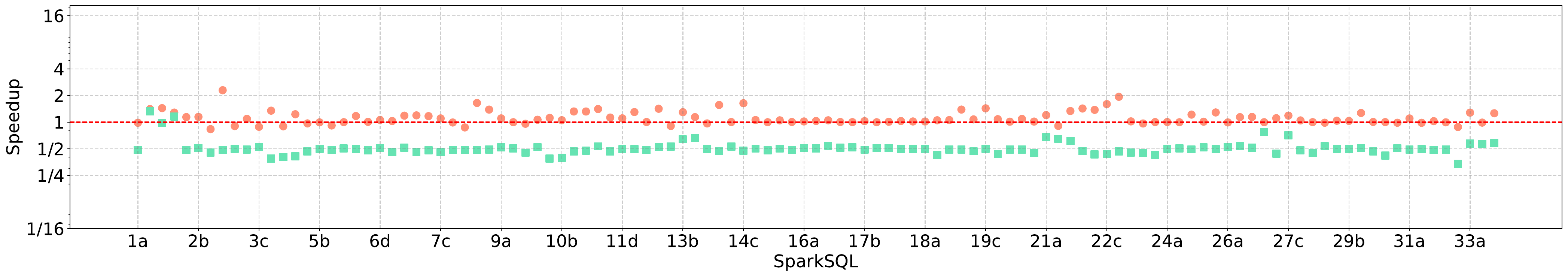}
        \label{fig:spark}
        \vspace*{-0.5cm}
    \end{minipage}
    \caption{Speedup achieved by different DBMS on JOB Benchmark.}
    \label{fig:job2}
\end{figure}

\begin{itemize}[leftmargin=*] \item \textbf{Sub-Graph Pattern Benchmark (SGPB).} Our rewriter significantly enhances performance across all systems on the SGPB benchmark. In DuckDB, we achieve a maximum speedup of 47,059x and an average of 194x over the native plans. AnalyticDB shows a maximum speedup of 6,606x with an average of 29x, PostgreSQL reaches up to 9,600x with an average of 107x, and SparkSQL records a maximum of 89x and an average of 2.7x. These results highlight the effectiveness of our rewrite algorithm, especially in analytical processing systems like DuckDB.

\item \textbf{LSQB (Scale Factor 30).} The rewriter provides substantial speedups on the LSQB benchmark. DuckDB experiences a maximum speedup of 2,391x and an average of 14x. AnalyticDB achieves up to 1,016x with an average of 9x, PostgreSQL sees a maximum of 67x and an average of 7x, while SparkSQL records a maximum of 538x with an average of 18x. Notably, several native query plans exceeded time limits or failed due to memory constraints; post-optimization, these queries were completed successfully, particularly Q8 and Q9. 

\item \textbf{TPC-H (Scale Factor 100).} Although the PK-FK constraints on the benchmark also limit the improvement of the new rewriting approach, it is still able to achieve some performance gains and avoid running time drawbacks by optimizing the number of rewritten queries and the query plan for PK-FK joins.  DuckDB shows a maximum speedup of 1.33x with an average of 1.06x. AnalyticDB reaches up to 3.93x with an average of 1.20x, PostgreSQL has a maximum of 1.75x and an average of 1.08x, and SparkSQL records a maximum of 1.09x with an average of 1.02x.  In addition, our rewrite query plan has at most $12.75\%$ performance drawbacks. 

\item \textbf{JOB.} The rewriter's performance on the Join Order Benchmark (JOB) is mixed. DuckDB achieves a maximum speedup of 14.84x with an average of 1.42x. AnalyticDB reaches up to 94.50x and averages 2.71x, PostgreSQL shows a maximum of 12.31x with an average of 1.40x, and SparkSQL records a maximum of 2.30x and an average of 1.11x.    In addition, to provide deeper insights, Table~\ref{tab:job} presents statistical analyses of the running times for all 113 queries in the JOB benchmark, which indicate significant improvements across various statistical measures.


\item \textbf{CEB. }Given the vast number of queries, we selected $5$ queries for testing. Notably, the experimental results for DuckDB reach a maximum 5.03x speedup and an average speedup of 3.33x. For PostgreSQL, the maximum speedup was 2.65x, and the average speedup was 1.54x.

\end{itemize}

\begin{table}[H]
\captionsetup{
    justification=centering, 
    singlelinecheck=false 
}
\centering
\begin{minipage}[t]{0.48\textwidth} 
    \centering
    \caption{JOB Statistics}
    \label{tab:job}
    \resizebox{\textwidth}{!}{ 
    \begin{tabular}{l|c|cccc}
    \toprule
    \textbf{Method (s)} & \textbf{Max} & \textbf{Mean} & \textbf{Med.} & \textbf{Std.Dev.} \\
    \midrule
    DuckDB native & 933.73 & 53.02 & 40.72 & 113.47 \\
    DuckDB Yannakakis & 262.67 & 45.08 & 44.42 & 33.17 \\
    DuckDB \name & 67.48 & 30.12 & 28.32 & 20.50 \\
    \midrule
    AnalyticDB native & 1282.22 & 106.13 & 59.78 & 194.39 \\
    AnalyticDB Yannakakis & 1468.3 & 226.19 & 175.51 & 220.85 \\
    AnalyticDB \name & 110.28 & 31.66 & 19.01 & 29.72 \\
    \midrule
    PostgreSQL native & 1289.29 & 82.66 & 56.36 & 147.97 \\
    PostgreSQL Yannakakis & 422.49 & 113.52 & 92.81 & 89.19 \\
    PostgreSQL \name & 144.16 & 50.55 & 43.56 & 35.68 \\
    \midrule
    SparkSQL native & 539.37 & 268.37 & 201.71 & 159.64 \\
    SparkSQL Yannakakis & 1145.17 & 544.72 & 430.47 & 328.92 \\
    SparkSQL \name & 521.33 & 207.56 & 170.81 & 156.95 \\
    \bottomrule
    \end{tabular}
    }
    \hspace{1em}
    \caption{Rule-based Optimization: PK-FK \& Annot}
    \label{tab:rule}
    \resizebox{\textwidth}{!}{ 
    \begin{tabular}{l|ccccc}
    \toprule
    \textbf{JOB-1a (s)} & \textbf{Base} & \textbf{Primitive} & \textbf{PK-FK} & \textbf{Annot} & \textbf{PK-FK \& Annot}\\
    \midrule
    DuckDB & 4.36 & 29.68 & 4.51 & 27.97 & 3.59\\
    \midrule
    PostgreSQL & 7.55 & 29.18 & 9.56 & 14.60 & 6.95\\
    \midrule
    \textbf{JOB-4a (s)} & \textbf{Base} & \textbf{Primitive} & \textbf{PK-FK} & \textbf{Annot} & \textbf{PK-FK \& Annot}\\
    \midrule
    DuckDB & 12.76 & 32.31 & 4.28 & 31.25 & 4.08\\
    \midrule
    PostgreSQL & 10.87 & 29.11 & 7.13 & 28.18 & 6.72\\
    \bottomrule
    \end{tabular}
    }
\end{minipage}
\hfill 
\begin{minipage}[t]{0.49\textwidth} 
    \centering
    \vspace{1em} 
    \caption{Running Times Under Different Cardinality Estimation Scenarios}
    \label{tab:cost}
    \resizebox{\textwidth}{!}{ 
    \begin{tabular}{l|ccccc}
    \toprule
    \textbf{JOB-2b (s)} & \textbf{native} & \textbf{accurate} & \textbf{estimated} & \textbf{worst-case bounds}\\
    \midrule
    DuckDB & 5.14 & 4.28 & 5.10 & 22.13\\
    \midrule
    PostgreSQL & 28.27 & 10.70 & 12.82 & 16.75\\
    \midrule
    \textbf{JOB-8b (s)} & \textbf{native} & \textbf{accurate} & \textbf{estimated} & \textbf{worst-case bounds}\\
    \midrule
    DuckDB & 23.60 & 22.74 & 23.38 & 38.00\\
    \midrule
    PostgreSQL & 92.19 & 59.86 & 85.97 & 97.32\\
    \midrule
    \textbf{JOB-11d (s)} & \textbf{native} & \textbf{accurate} & \textbf{estimated} & \textbf{worst-case bounds}\\
    \midrule
    DuckDB & 58.58 & 5.42 & 7.77 & 228.21\\
    \midrule
    PostgreSQL & 20.06 & 7.26 & 10.91 & 50.10\\
    \midrule
    \textbf{JOB-17c (s)} & \textbf{native} & \textbf{accurate} & \textbf{estimated} & \textbf{worst-case bounds}\\
    \midrule
    DuckDB & 39.20 & 16.24 & 20.46 & 35.90\\
    \midrule
    PostgreSQL & 72.45 & 69.73 & 70.30 & 377.29\\
    \midrule
    \textbf{JOB-27b (s)} & \textbf{native} & \textbf{accurate} & \textbf{estimated} & \textbf{worst-case bounds}\\
    \midrule
    DuckDB & 41.49 & 40.46 & 41.40 & 53.81\\
    \midrule
    PostgreSQL & 38.85 & 21.72 & 38.30 & 79.3 \\
    \bottomrule
    \end{tabular}
    }
\end{minipage}
\end{table}

\subsubsection{\textbf{Effectiveness of the Rule-based Optimization. }} 
We conducted ablation experiments to test the effects of two rules: PK-FK projection elimination and pruning for annotation. We select 1a and 4a query from the JOB benchmark, where \textit{base} represents the effect without any rewrite, \textit{primitive} represents the result without both rewrite rules, \textit{PK-FK} represents the effect with only projection elimination, \textit{Annot} represents the effect with only pruning for annotation, and \textit{PK-FK \& Annot} represents the combined effect of both optimizations. We test the experimental performance under two DBMSs and find that applying both optimizations simultaneously yields excellent experimental results, as shown in Table \ref{tab:rule}.



\subsubsection{\textbf{Effectiveness of Cardinality Estimation. }} To test our cost-based optimizer, we evaluate the impact of cardinality estimation accuracy on query performance under three scenarios:

\begin{itemize}[leftmargin=*]
\item \textbf{Accurate Cardinality}: The optimizer uses exact sizes for all intermediate query results. 
\item \textbf{Estimated Cardinality}: The optimizer relies on estimates based on available statistics like cardinalities and the number of distinct values (NDV). 
\item \textbf{Worst-Case Bounds}: The optimizer assumes maximum possible join sizes (Cartesian product) unless key constraints are present. 
\end{itemize}

Table~\ref{tab:cost} presents the execution times for three queries on DuckDB and PostgreSQL under these scenarios, along with the native plans.  The results indicate that accurate cardinality leads to optimal performance, while with estimated statistics, execution times improve significantly over the native plans and can provide similar performance compared with the optimal estimation.  On the other hand, we also need some accuracy to ensure the performance, as if we only apply the worst-case estimation, the performance can be much worse than our current selection or even native plans.

\begin{figure*}[h]
    \centering
    \begin{minipage}[t]{\textwidth}
        \hspace*{1.2cm}
        \includegraphics[width=0.9\textwidth]{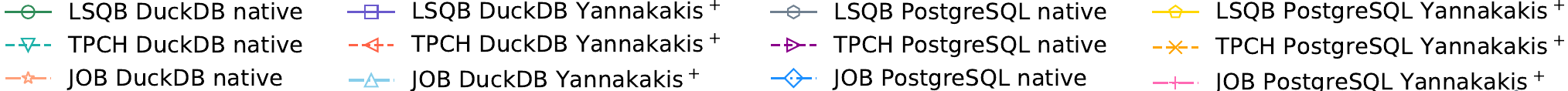}
        \label{fig:combined_tag}
    \end{minipage}
    \vspace{-0.2cm}  
    \subfigure[Selectivity]{
        \begin{minipage}[t]{0.49\textwidth}
        \centering
        \includegraphics[width=\textwidth]{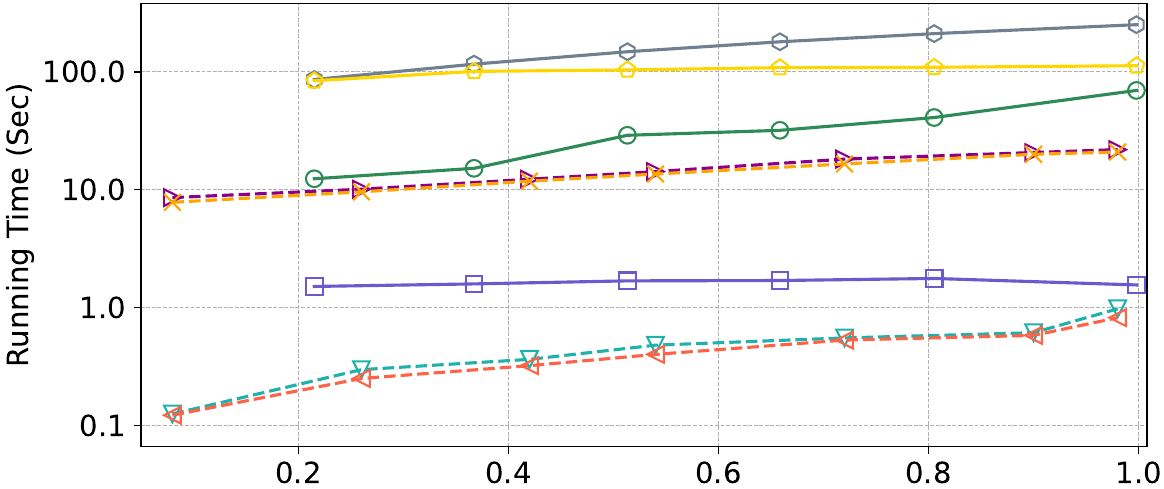}
        \label{fig:selectivity}
        \end{minipage}
    }
    \hfill  
    \subfigure[Scale]{
        \begin{minipage}[t]{0.47\textwidth}
        \centering
        \includegraphics[width=\textwidth]{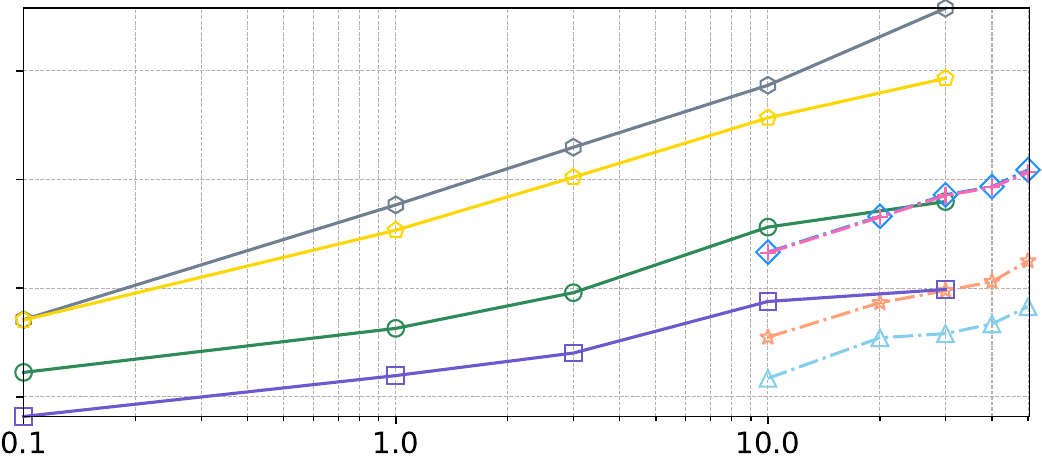}
        \label{fig:scale}
        \end{minipage}
    }
    \caption{Running times of different selectivity \& scale. }
\end{figure*}

\begin{figure*}[htbp]
\centering 
\subfigure[LSQB-Q1]{
    \includegraphics[width=0.48\textwidth]{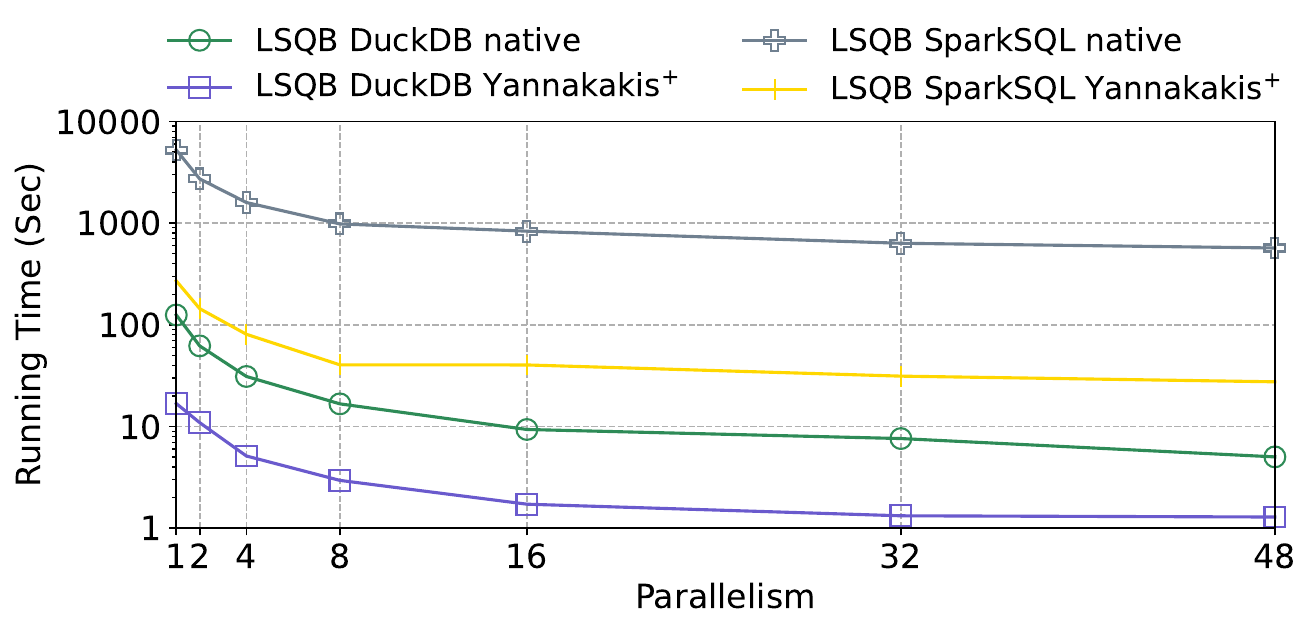} 
    \label{fig:thread} 
}
\hfill
\subfigure[SGPB-Q1]{
    \includegraphics[width=0.48\textwidth]{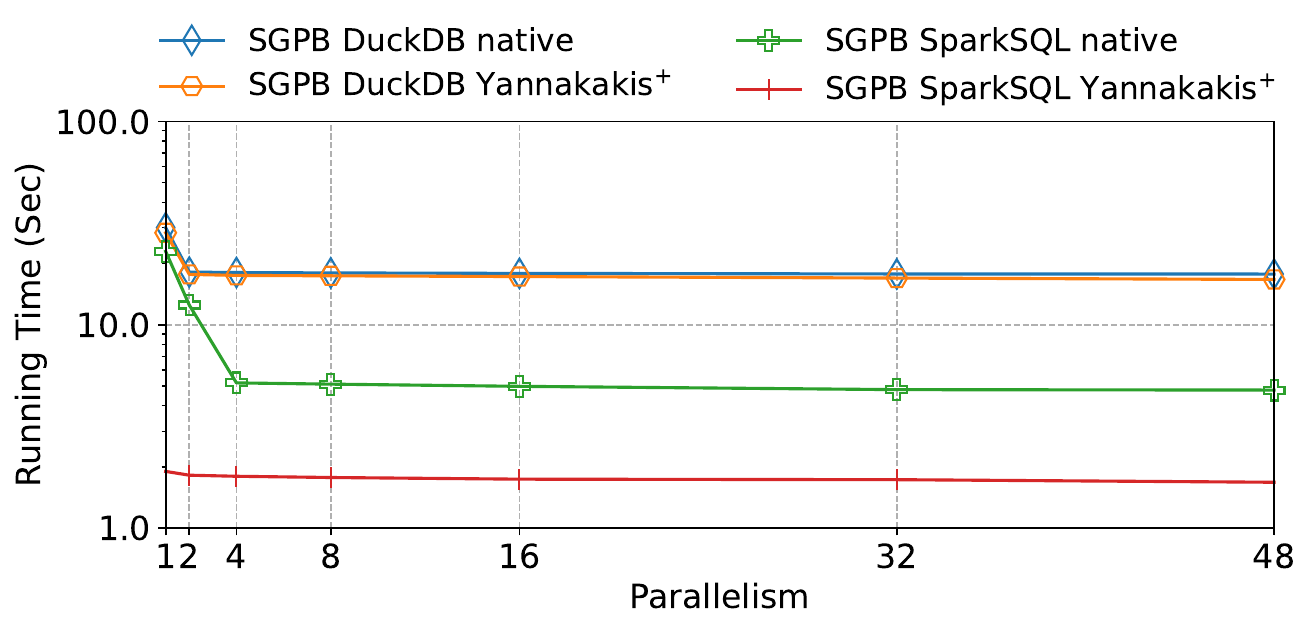} 
    \label{fig:thread2}
}
\vspace{-0.2cm}
\caption{Running times under different parallelism.}
\label{fig:parallelism}
\end{figure*}

\begin{table*}[htbp]
\centering
\caption{Optimizaiton time for different queries. }
\resizebox{\textwidth}{!}{ 
\begin{tabular}{l|cccccccc}
\toprule
\textbf{Query (s)} & \textbf{DuckDB native} & \textbf{DuckDB \name} & \textbf{PostgreSQL native} & \textbf{PostgreSQL \name} & \textbf{\#Tables} & \textbf{\#Attributes} & \textbf{Opt-Time} & \textbf{DuckDB Opt-Time}\\
\midrule
\textbf{SNAP-q1a} & 15.10 & 8.19 & 46.96 & 37.64 & 3 & 6 & 0.133703232 & 0.0021\\
\midrule
\textbf{SNAP-q6} & 8.12 & 2.29 & 146.01 & 46.69 & 5 & 6 & 0.236415863 & 0.0014\\
\midrule
\textbf{LSQB-q1} & 6.27 & 0.97 & 376.31 & 85.34 & 10 & 7 & 0.065845966 & 0.0087\\
\midrule
\textbf{LSQB-q5} & 10.37 & 7.47 & 153.74 & 151.53 & 3 & 4 & 0.087508917 & 0.0017\\
\midrule
\textbf{TPCH-q3} & 5.32 & 5.07 & 79.99 & 68.39 & 3 & 8 & 0.071694851 & 0.0019\\
\midrule
\textbf{TPCH-q10} & 12.36 & 9.32 & 33.25 & 32.24 & 4 & 13 & 0.085761070 & 0.0027\\
\midrule
\textbf{TPCH-q19} & 5.72 & 5.68 & 52.94 & 51.89 & 2 & 9 & 0.074432135 & 0.0020\\
\midrule
\textbf{JOB-1a} & 3.66 & 3.21 & 7.34 & 6.72 & 5 & 8 & 0.075666189 & 0.0027\\
\midrule
\textbf{JOB-10c} & 23.59 & 23.49 & 131.33 & 92.96 & 7 & 10 & 0.172396183 & 0.0051\\
\midrule
\textbf{JOB-21a} & 40.93 & 40.01 & 56.36 & 36.78 & 9 & 13 & 0.080693007 & 0.0137\\
\midrule
\textbf{JOB-27c} & 41.10 & 40.76 & 51.22 & 36.65 & 12 & 17 & 0.086112976 & 0.0594\\
\midrule
\textbf{JOB-30a} & 61.14 & 35.86 & 60.24 & 49.86 & 14 & 21 & 0.096741199 & 0.0666\\
\bottomrule
\end{tabular}
}
\label{tab:opt}
\end{table*}

\subsubsection{\textbf{Robustness.} }  From the experimental results, \name consistently shows improvements in the vast majority of queries tested and shows excellent robustness.  This is mainly benefiting from \name's optimization of the number of semi-joins.  In most queries, only one round or even no semi-join reduction is required.  The rule-based optimizer also helps avoid unnecessary semi-join reductions.  Beyond semi-join reduction, aggregation pushdown also contributes to improved performance.  Furthermore, we also tested \name under different selectivity and data scales to further illustrate its robustness along these dimensions:

\noindent \textbf{Selectivity.} We selected two queries and altered the predicate to change $F$ (the full join size). In Figure \ref{fig:selectivity}, the horizontal axis values represent the percentage of output value (size) compared with output value (size) without a predicate. It can be observed that as the output size increases, the advantages of the rewriter increase compared to the original query execution. 

\noindent \textbf{Scale.} We selected five settings for LSQB: 0.1, 1, 3, 10, and 30. For JOB, we chose five scales with equal intervals from 10 to 50. The curves in Figure \ref{fig:scale} show that as the scale increases, the running time generally increases proportionally, and the larger the scale, the better the execution performance of our rewrite.

\noindent \textbf{Parallel Query Processing.}
Our approach is pure relational, which offers the significant benefit of seamless adoption in other computing scenarios, such as parallel processing. We conducted an experiment where we varied the number of threads utilized by each DBMS and re-executed a set of specific queries, with LSQB-Q1 selected in Figure \ref{fig:thread} and SGPB-Q1 in Figure \ref{fig:thread2}. From the experimental results, our new query plan also shows improvement with additional threads, which is similar to that of the native query plan, indicating great parallelization of our new plan. 

\noindent \textbf{Optimization Time. } Finally, it is worth mentioning that the optimization overhead is relatively small compared to the total query execution time.
We selected representative 12 queries from the four benchmarks to investigate the relationship between optimization time and the number of tables and attributes within the query.   As seen from Table \ref{tab:opt}, the optimization time is mostly kept within 100ms, which is negligible compared to the query execution time.  This favorable outcome is partially attributed to the introduction of a hint mechanism within our system.  When the optimization time reaches a certain threshold, we leverage the existing plans from the DBMS to assist our estimator.  We do observe that there is a performance gap between the optimization time of \name and that of DuckDB.  However, this gap is well offset by the performance gains in the query execution time.  Furthermore, we believe this gap can be significantly reduced if we integrate \name within the database kernel---recall that the current implementation of our optimizer is outside the engine, thus incurring quite some overhead (in exchange for better compatibility with different engines). 




\section{Conclusion and Future Work}
\label{sec:future}
In this work, we introduce \name, an improved version of the original Yannakakis algorithm. This new version not only maintains the theoretical guarantees but is also highly efficient in practice.  The experimental results suggest that \name can not only achieve order-of-latitude improvements on specific queries while avoiding regressions on other queries.  

\revise{Our current implementation of \name follows a rewrite-based approach to showcase its applicability across a wide range of database and data processing systems (row-based vs column-based, centralized vs distributed).  The next natural step is to integrate it into an SQL engine, which could further improve its performance.  For example, we can generate a single physical plan instead of issuing multiple SQL statements, reduce the communication overhead between system components, and eliminate repeated parsing, plan generation, and optimization. Beyond these direct advantages, combining \name with the database engine offers further opportunities for optimization: (1) Note that our use of semijoin is ``soft'', i.e., it is alright to leave a small number of dangling tuples unremoved.  So this can be using Bloom filters, which are much more efficient than using the existing semi-join operator. (2) Implementing \name inside database engines enables access to more sophisticated database statistics, potentially improving the cost-based optimizer with tailored-made CE/CM. (3) The bottleneck of the optimization time in \name is running GYO and GHD to enumerate all possible query plans. By integrating \name within database engines, it is possible to reduce redundant plan enumeration through the native database optimizer, which is especially useful for large queries with hundreds of relations or attributes. These enhancements are expected to boost \name's performance and make it a more robust solution for advanced query processing scenarios.}


\bibliographystyle{ACM-Reference-Format}
\bibliography{ref}
\appendix

\onecolumn

\section{Missing Proofs in Section 2}

\subsection{Running example for Example 2.1}
The figure illustrates the execution diagram for Example \ref{ex:query}. Each arrow corresponds to a set of tables with the same color on the left side, and the table on the right side of the arrow represents the result of their project \& join operations. The arrow itself signifies a set of rewrite queries involving project \& join. The colored columns on the tables indicate the columns to be projected, and the colored rows within the tables represent the tuples that can be joined together. TempView denotes the intermediate results that emerge during the rewrite process. The entire process flows from left to right, with the final result displayed in the Result table.
\begin{figure*}[ht]
\centering
\resizebox{\linewidth}{!}{
\begin{tikzpicture}

    \node (R3) at (-4, 0.7) [scale=0.31] {
        \begin{tabular}{|c|c|c|}
            \multicolumn{3}{c}{$\textbf{R\textsubscript{3}}$}\\ \hline
            \cellcolor{red!30}$x_3$ & \cellcolor{red!30}$x_4$ & \cellcolor{red!30}annot\\ \hline
            \cellcolor{red!30}1 & \cellcolor{red!30}1 & \cellcolor{red!30}6\\
            \cellcolor{red!30}1 & \cellcolor{red!30}2 & \cellcolor{red!30}7\\
            \cellcolor{red!30}2 & \cellcolor{red!30}2 & \cellcolor{red!30}8\\\hline
        \end{tabular}
    };

    \node (R4) at (-4,0) [scale=0.31] {
        \begin{tabular}{|c|c|c|}
            \multicolumn{3}{c}{$\textbf{R\textsubscript{4}}$}\\ \hline
            \cellcolor{red!30}$x_3$ & $x_6$ & \cellcolor{red!30}annot\\ \hline
            \cellcolor{red!30}1 & blue... & \cellcolor{red!30}1\\
            \cellcolor{red!30}2 & blue... & \cellcolor{red!30}1\\
            3 & blue... & 1\\ \hline
        \end{tabular}
    };

     \node (R5) at (-4,-0.7) [scale=0.31] {
        \begin{tabular}{|c|c|c|}
            \multicolumn{3}{c}{$\textbf{R\textsubscript{5}}$}\\ \hline
            \cellcolor{blue!30}$x_4$ & $x_7$ & \cellcolor{blue!30}annot\\ \hline
            \cellcolor{blue!30}1 & 1 & \cellcolor{blue!30}1\\
            \cellcolor{blue!30}2 & 2 & \cellcolor{blue!30}1\\
            \cellcolor{blue!30}3 & 1 & \cellcolor{blue!30}1\\ \hline
        \end{tabular}
    };

     \node (R6) at (-4,-1.4) [scale=0.31] {
        \begin{tabular}{|c|c|c|}
            \multicolumn{3}{c}{$\textbf{R\textsubscript{6}}$}\\ \hline
            $x_7$ & \cellcolor{blue!30}$x_8$ & \cellcolor{blue!30}annot\\ \hline
            1 & \cellcolor{blue!30}ARGENTINA & \cellcolor{blue!30}1\\
            2 & \cellcolor{blue!30}BRAZIL & \cellcolor{blue!30}1\\
            3 & CANADA & 1\\ \hline
        \end{tabular}
    };

    \node (temp1) at (-2.7,0.3) [scale=0.31] {
        \begin{tabular}{|c|c|c|}
            \multicolumn{3}{c}{$\textbf{TempView1}$}\\ \hline
            \cellcolor{brown!40}$x_3$ & \cellcolor{brown!40}$x_4$ & \cellcolor{brown!40}annot\\ \hline
            \cellcolor{brown!40}1 & \cellcolor{brown!40}1 & \cellcolor{brown!40}6\\
            1 & 2 & 7\\
            \cellcolor{brown!40}2 & \cellcolor{brown!40}2 & \cellcolor{brown!40}8\\ \hline
        \end{tabular}
    };

    \node (R1) at (-2.7, -0.45) [scale=0.31] {
        \begin{tabular}{|c|c|c|c|c|}
            \multicolumn{5}{c}{$\textbf{R\textsubscript{1}}$}\\ \hline
            \cellcolor{brown!40}$x_1$ & \cellcolor{brown!40}$x_2$ & $x_3$ & \cellcolor{brown!40}$x_4$ & \cellcolor{brown!40}annot \\ \hline
            \cellcolor{brown!40}4 & \cellcolor{brown!40}1 & 1 & \cellcolor{brown!40}1 & \cellcolor{brown!40}3\\
            5 & 1 & 2 & 1 & 4\\
            \cellcolor{brown!40}6 & \cellcolor{brown!40}1 & 2 & \cellcolor{brown!40}2 & \cellcolor{brown!40}5\\\hline
        \end{tabular}
    };

    \node (temp2) at (-1.3,-1.1) [scale=0.31] {
        \begin{tabular}{|c|c|c|}
            \multicolumn{3}{c}{$\textbf{TempView2}$}\\ \hline
            $x_4$ & \cellcolor{red!30}$x_8$ & \cellcolor{red!30}annot\\ \hline
            1 & \cellcolor{red!30}ARGENTINA & \cellcolor{red!30}1\\
            2 & \cellcolor{red!30}BRAZIL & \cellcolor{red!30}1\\
            3 & ARGENTINA & 1\\ \hline
        \end{tabular}
    };

    \node (temp3) at (-1.4,-0.1) [scale=0.31] {
        \begin{tabular}{|c|c|c|c|}
            \multicolumn{4}{c}{$\textbf{TempView3}$}\\ \hline
            \cellcolor{red!30}$x_1$ & \cellcolor{red!30}$x_2$ & $x_4$ & \cellcolor{red!30}annot\\ \hline
            \cellcolor{red!30}4 & \cellcolor{red!30}1 & 1 & \cellcolor{red!30}18\\
            \cellcolor{red!30}6 & \cellcolor{red!30}1 & 2 & \cellcolor{red!30}40\\ \hline
        \end{tabular}
    };

    \node (R2) at (0.2, -1.3) [scale=0.31] {
        \begin{tabular}{|c|c|c|}
            \multicolumn{3}{c}{$\textbf{R\textsubscript{2}}$}\\ \hline
            \cellcolor{blue!30}$x_2$ & $x_5$ & \cellcolor{blue!30}annot \\ \hline
            \cellcolor{blue!30}1 & 1996 & \cellcolor{blue!30}1\\
            2 & 1996 & 1\\
            3 & 1996 & 1\\\hline
        \end{tabular}
    };

    \node (temp4) at (0.2,-0.5) [scale=0.31] {
        \begin{tabular}{|c|c|c|c|}
            \multicolumn{4}{c}{$\textbf{TempView4}$}\\ \hline
            \cellcolor{blue!30}$x_1$ & \cellcolor{blue!30}$x_2$ & \cellcolor{blue!30}$x_8$ & \cellcolor{blue!30}annot\\ \hline
            \cellcolor{blue!30}4 & \cellcolor{blue!30}1 & \cellcolor{blue!30}ARGENTINA & \cellcolor{blue!30}18\\
            \cellcolor{blue!30}6 & \cellcolor{blue!30}1 & \cellcolor{blue!30}BRAZIL & \cellcolor{blue!30}40\\ \hline
        \end{tabular}
    };

     \node (res) at (1.9,-1) [scale=0.31] {
        \begin{tabular}{|c|c|c|c|}
            \multicolumn{4}{c}{$\textbf{Result}$}\\ \hline
            $x_1$ & $x_2$ & $x_8$ & annot\\ \hline
            4 & 1 & ARGENTINA & 18\\
            6 & 1 & BRAZIL & 40\\ \hline
        \end{tabular}
    };

    \node at (-3.5, 0.22) [scale=0.3, above] {Project \& Join};
    \draw[->] (-3.8, 0.25) -- (-3.1, 0.25);
    
    \node at (-2.7, -1.1) [scale=0.3, above] {Project \& Join};
    \draw[->] (-3.5, -1.1) -- (-1.95, -1.1);
    
    \node at (-2.29, -0.2) [scale=0.3, above] {Project \& Join};
    \draw[->] (-2.55, -0.2) -- (-1.9, -0.2);
    
    \node at (-0.92, -0.6) [scale=0.3, above] {Project \& Join};
    \draw[->] (-1.2, -0.6) -- (-0.53, -0.6);

    \node at (0.75, -1) [scale=0.3, above] {Project \& Join};
    \draw[->] (0.4, -1) -- (1.15, -1);


\end{tikzpicture}
}
\caption{ A running example of Example \ref{ex:query}}
\label{fig:rewriteAgg}
\end{figure*}
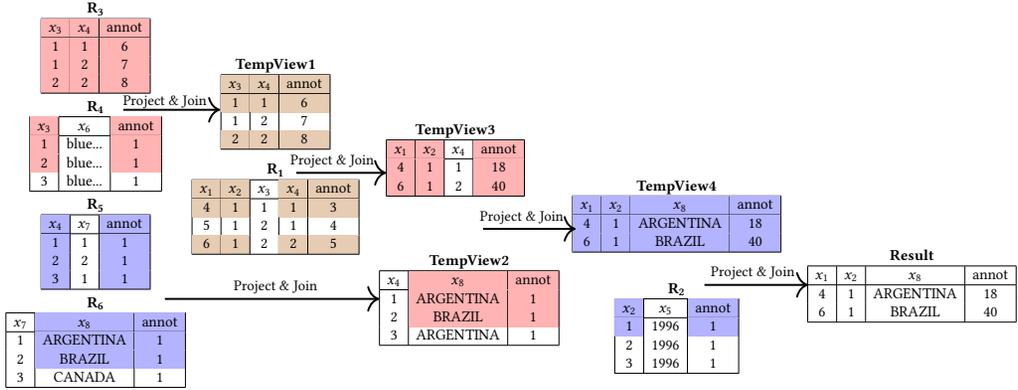

\subsection{Missing proof for Lemma~\ref{lem:free-connex}}

To create a join tree $\T'$ for $\Q \Join [\bm{O}]$ using the given join tree $\T$, follow these steps:
\begin{enumerate}
\item Set $[\bm{O}]$ as the root of $\T'$.
\item Connect all relations in $\T_n$ to be the child nodes of $[\bm{O}]$.
\item For the remaining relations in $\T \setminus \T_n$, connect them on $\T'$ to the same parent node as in $\T$.
\end{enumerate}
By following these steps, it is evident that $\T'$ is a valid join tree for $\Q \Join [\bm{O}]$, ensuring it is acyclic.

On the other hand, since both $\Q$ and $\Q \Join [\bm{O}]$ are acyclic, a join tree $\T'$ for $\Q \Join [\bm{O}]$ with $[\bm{O}]$ as the root node can be constructed by initially building an arbitrary join tree and then rotating the tree until $[\bm{O}]$ is the root node.  Let $\C_r$ be all the child nodes of $[\bm{O}]$ on $\T'$, all nodes in $\C_r$ must satisfy 
\begin{itemize}
    \item $\bm{O} \subseteq \bigcup_{R(\bm{A}) \in \C_r} \bm{A}$; and
    \item $\forall R_i(\bm{A}_i), R_j(\bm{A}_j) \in \C_r, \bm{A}_i \cap \bm{A}_j \subseteq \bm{O}$.
\end{itemize}
Otherwise, the connected property for attributes is broken on the tree $\T'$.  If all relations in $\C_r$ is acyclic, then a free-connex join tree $\T$ can be constructed by 
\begin{enumerate}
    \item Creating a join tree on all relations in $\C_r$; such a join tree will be the connex subset of $\T$ because the two properties for $\C_r$; and
    \item For the remaining relations in $\bm{R} \setminus \C_r$, connect them on $\T$ to the same parent node as in $\T'$.
\end{enumerate}
The crucial factor is whether $\C_r$ is acyclic. It is possible that for an acyclic query, the query induced by a subset of its relations is not acyclic. For instance, $R_1(x_1, x_2, x_3) \Join R_2(x_1, x_2) \Join R_3(x_2, x_3) \Join R_4(x_1, x_3)$ is acyclic, while $R_2(x_1, x_2) \Join R_3(x_2, x_3) \Join R_4(x_1, x_3)$ is cyclic. If we assume that a set of relations $C \in \C_r$ forms a cycle, and since $\Q$ is acyclic, there exists a relation $R \in \bm{R}$ such that for every pair of $R_i, R_j \in C$, $R_i \cap R_j \subseteq R$. Furthermore, $R \in \C_r$; otherwise, the connected property is broken on $\T'$. When we combine all this, we can conclude that all cycles in $\C_r$ have an $R$ that contains all join attributes in $\C_r$, making the cycle reducible by $R$. Hence, $\C_r$ is acyclic, which completes the proof.

\section{Missing Materials for Section~3}

\begin{figure*}[h]
\centering
        \subfigure[Standard query plan.]
        {
            \begin{minipage}[t]{0.2\linewidth}
            \centering
            \adjincludegraphics[width=\linewidth]{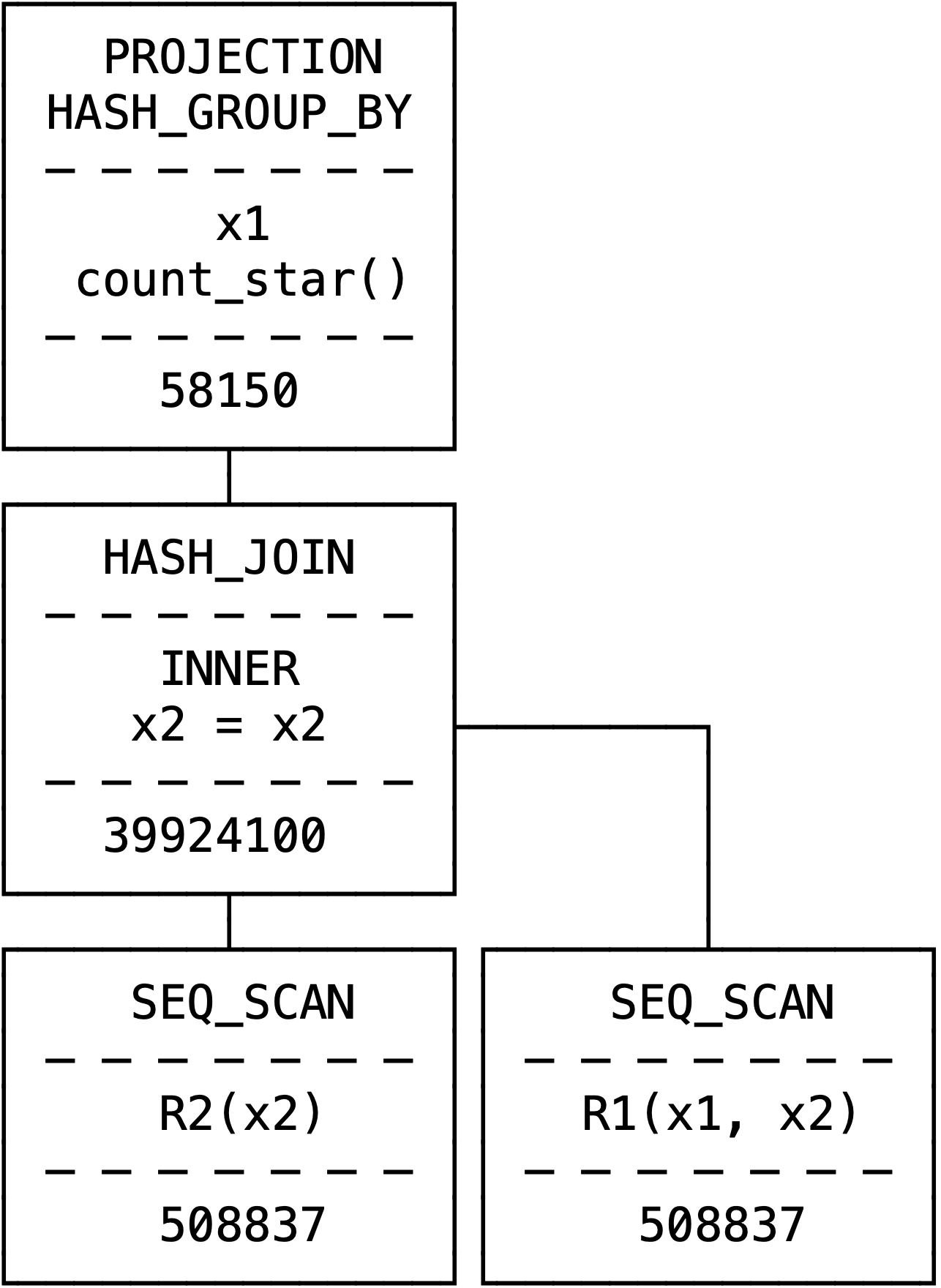}
            \label{fig:binary_original}
            \end{minipage}
        }
        \hspace{3em}
        \subfigure[New query plan.]
        {
            \begin{minipage}[t]{0.2\linewidth}
            \centering
            \adjincludegraphics[width=\linewidth]{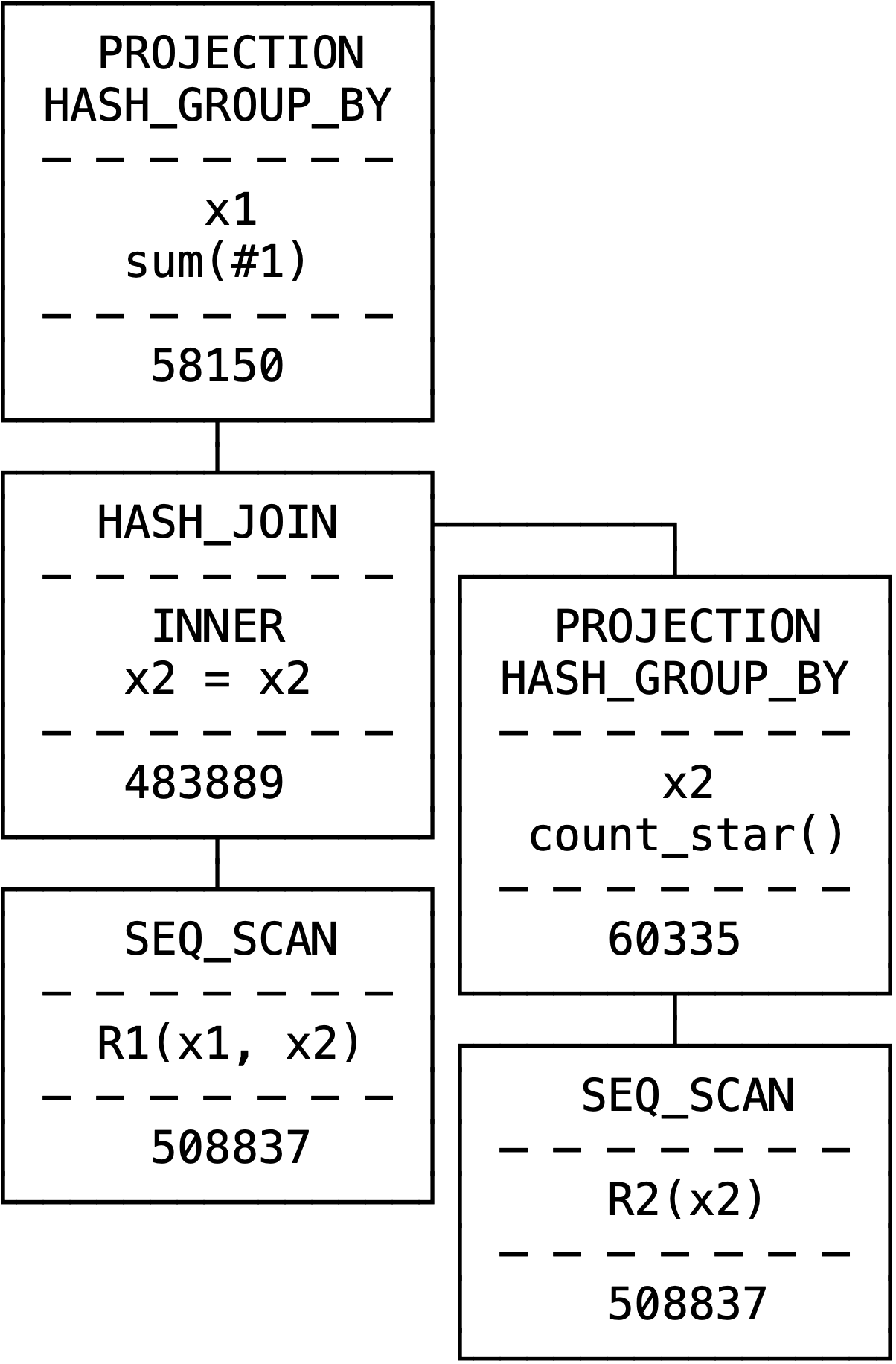}
            \label{fig:binary_rewrite}
            \end{minipage}
        }
        \hspace{3em}
        \subfigure[Yannakakis query plan.]
        {
            \begin{minipage}[t]{0.3\linewidth}
            \centering
            \adjincludegraphics[width=\linewidth]{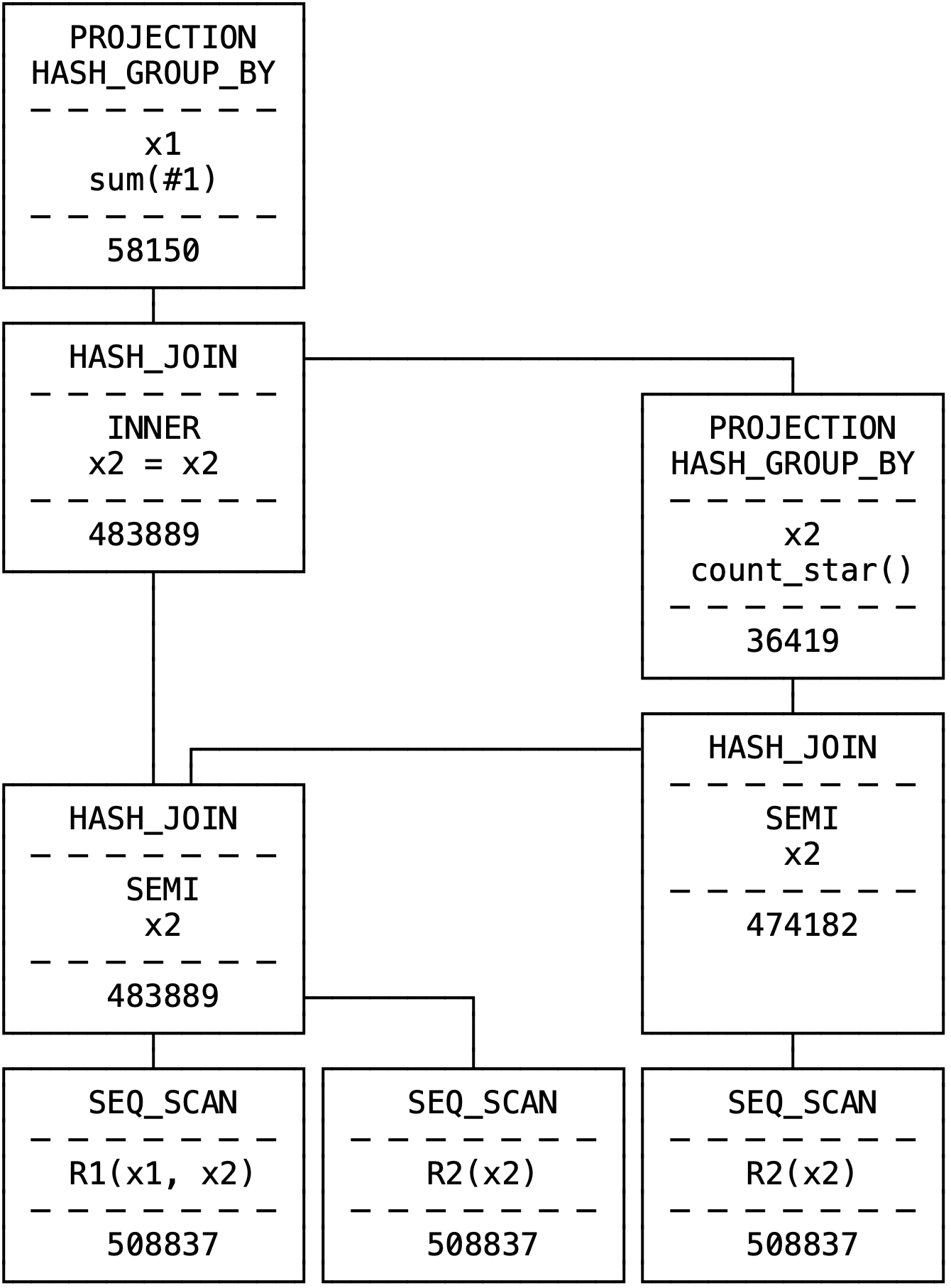}
            \label{fig:binary_yannakakis}
            \end{minipage}
        }
        \caption{Query plans for $\Q_4$ in DuckDB. }
        \label{fig:binary}
\end{figure*}

\subsection{Missing Proof for Lemma~\ref{lem:3.6}}

\begin{proof}
We prove inductively that each step in Algorithm \ref{alg:R1} maintains the equivalence.  First, the semi-join in line 9 does not change the query results, since it only removes dangling tuples while leaving all the annotations untouched. Second, the aggregation in line 8 aggregates out only non-output attributes that uniquely appear in $R_i$, and equivalence follows from the distributive law of the semiring.  

It remains to argue for the correctness of lines 5--6. If the leaf node $R_i$ does not contain any unique output attributes, lines 5--6 perform the join between $\left(\pi_{\A_p} R_i\right)$ and its parent node $R_p$, and remove $R_i$ from $\Q$, i.e., the reduced query is
    \[
        \Q' := \pi_{\O} \left(\left(\Join_{k \in [n], k \neq i} R_k\right) \Join \left(\pi_{\A_p} R_i\right) \right).
    \]
    For any $t \in \Q$, there exists a tuple $t' \in \mathcal{J}$ such that $\pi_{\O} t' = t$.  Since $t' \in \mathcal{J}$, $\pi_{\bar{\A}_i} t' \in \left(\Join_{k \in [n], k \neq i} R_k\right)$ and $\pi_{\A_i} t' \in R_i$, making $\pi_{\bar{\A}_i} t' \in \left(\left(\Join_{k \in [n], k \neq i} R_k\right) \Join \left(\pi_{\A_p} R_i\right) \right)$.  Therefore, since $R_i$ contains no unique output attribute, $O \cap \bar{\A}_i = \O$, $\pi_{\bar{\A}_i} t' = t$.  
\end{proof}

\subsection{Missing Proof for Lemma~\ref{lem: output-only}}

\begin{proof}

The algorithm removes all unique non-output attributes at Line~$8$ in Algorithm~\ref{alg:R1}, making the $\Q'$ contain only output attributes and join attributes and satisfy (1) for non-free-connex queries.

If $\Q$ is free-connex, we can further show the join attributes between the non-root node $R_i$ and its parent node $R_p$ as output attributes, i.e., $\A_i \cap \A_p \subseteq \O$, after removing all unique non-output attributes. To prove that, let's assume there are some non-output join attributes after the first-round computation for the free-connex query. Let $R_i$ be a non-root relation such that $\A_i \cap \A_p \nsubseteq \O$ and $\A_p \cap \O \neq \emptyset$. It is necessary for such $R_i$ to exist. If not, then for all $\A_i$ that contains non-output attributes, $\A_p$ has no output attribute, and we can further deduce that for the root node $R_r$, $\A_r$ has no output attribute. Therefore, $\O$ must be $\emptyset$ to ensure that $\T'$ is a free-connex join tree, and the resulting join tree would contain only a single relation.

On the other hand, if such $R_i$ exists, there won't be any output attributes $x \in \O$, $x \notin \A_p$ on the subtree rooted on $R_i$. If there were, the free-connex condition would be violated as that node needs to be contained in the connex subset. This means that the subtree rooted on $R_i$ would be eliminated from $\T'$ during the post-order traversal on $\T$, since $(\A_i - \A_p) \cap \O$ holds for every node on the subtree, which contradicts the assumption of $R_i \in \T'$ and proves~(2).

From (2), we can further prove (3) as there can be at most one relation $R$ containing output attributes for a relation-dominated $\Q$, and the resulting $\Q'$ is a full query, indicating that only $R$ can appear in $\Q'$.
\end{proof}

\subsection{Missing Proof for Lemma~\ref{lem:free-root}}

\begin{proof}
The lemma can be proven by inducting the structure of the join tree. For the sake of clarity, let's denote $R_i$ as the relation after the first-round computation and $R_i'$ as the original relation before the first-round computation. If $R_i$ was removed during the first-round computation, we conceptually set $R_i$ as the relation before deleting from $\T$.  We define the subquery $\Q_i$ on the subtree $\T_i$ of $\T$, rooted on $R_i$, as $\Q_i=\Join_{k \in \T'_i} R'_k$. After the first-round computation, we can demonstrate that $R_i=R'_i \ltimes \Q_i$ for any $R_i \in \T$ by considering the following conditions:
\begin{itemize}[leftmargin=*]
\item \textbf{$R_i$ is a leaf node.} The condition obviously holds since $\Q_i = R_i'$;
\item \textbf{$R_i$ is a non-leaf node.} If the condition holds for any $R_c \in \C_i$, and the relation $R_i$ satisfies
\[
    R_i = R'_i \ltimes_{R_c \in \C_i} R'_c
\]
ensuring by the semi-join in Line 9 and the join in Line 5 of Algorithm~\ref{alg:R1}.  
Thus, 
\[
\begin{aligned}
    R_i &= R'_i \ltimes_{R_c \in \C_i} R'_c \\
    &= R'_i \ltimes_{R_c \in \C_i} \Q_c \\
    &= R'_i \ltimes \Q_i
\end{aligned}
\]
\end{itemize}
Applying induction from the leaf nodes to the root nodes can show the condition holds for any tree nodes. Therefore, $R_r = R_r' \ltimes \Q_r$, and $\Q_r = \Q$ by definition. This means that all tuples in $R_r$ have at least one corresponding full join tuple, ensuring that $R_r$ is dangling-free.
\end{proof}

\subsection{Missing Proof for Lemma~\ref{lem:danglingbound}}

\begin{proof}
   Because the relation $R_i$ is dangling-free, and $R_j$ is its reducible relation, for every tuple $t \in R_i$, there exists a tuple $t' \in \mathcal{J}$ from the full join query such that $t = \pi_{\A_i} t'$. Additionally, the remaining attributes $\A_i - \A_j$ satisfy $\A_i - \A_j \subseteq \O$, due to the definition of a reducible relation and the fact that all unique non-output attributes are removed in the first-round computation.  This implies that $|\pi_{\A_j} \left( R_i \Join R_j\right)| \le |R_j| = O(N)$ and $|\pi_{\A_i - \A_j} \left(R_i \Join R_j\right)| \le \OUT$, making the total size of $R_i \Join R_j$ be bounded by $N\OUT$ and $F$ and the evaluation can be done in $O(\min(N\OUT, F))$.

   In addition, for any join result $t \in R_i \Join R_j$, there exists $t' \in \mathcal{J}$ such that $\pi_{\A_i \cup \A_j} t' = t$.  Hence, if $\bar{\A}_j \cap \A_j \subseteq \O$, then all attributes in $\A_i \cup \A_j$ are output attributes. For every $t \in R_i \Join R_j$, there exists at least one $t' \in \mathcal{J}$ with the corresponding $t'' \in \Q$ such that $pi_{\O} t' = t''$ and $\pi_{\A_i \cup \A_j} t' = t$, hence for every such $t$, $\pi_{\A_i \cup \A_j} t'' = t$ and there exists at least one such output tuples of $\Q$, limiting the total number of such $t$ to be at most $O(\OUT)$.
\end{proof}

\subsection{Missing Proof for Lemma~\ref{lem:childdangling}}

\begin{proof}
We divide all relations in $\mathcal{R}$ into two parts, $\mathcal{R}_j$ and $\bar{\mathcal{R}}_j$, where $\mathcal{R}_j$ represents all relations in the subtree $\T_j$ of $\T$, rooted at $R_j$, and $\bar{\mathcal{R}}_j$ represents all the remaining relations $\mathcal{R} \setminus \mathcal{R}_j$.  Let $\bar{\Q}_j := \Join_{R \in \bar{\mathcal{R}}_j} R$ and $\Q_j := \Join_{R \in \mathcal{R}_j} R$ After the first-round computation, $R_j$ satisfies for every $t_j \in R_j$, there exists a $t \in \Q_j$, such that $\pi_{\A_j} t = t_j$.

On the other hand, since $R_i$ is dangling-free, for any $t_i \in R_i$, there exists a $t \in \bar{R}_j$, such that $\pi_{\A_i} t = t_i$.  Therefore, for all $t_j \in R_j \ltimes R_i$, let $t_i \in R_i$ be one of the tuples that satisfies $\pi_{\A_i \cap \A_j} t_i = \pi_{\A_i \cap \A_j} t_j$, $t \in \bar{\Q}_j$ be one tuple that can join with $t_i$, and $t' \in \Q_j$ be one tuple that can join with $t_j$, then
\[
    t_i \Join t_j \Join t \Join t' \in \Q_j \Join \bar{\Q}_j,
\]
where $\Q_j \Join \bar{\Q}_j = \mathcal{J}$.  Therefore, for all $t_j \in R_j'$, we can find at least one corresponding full join result, making $R_j'$ dangling-free.
\end{proof}

\subsection{Missing Proof for Theorem~\ref{thm:non-free-connex}}

\begin{proof}
    Since the cost of any semi-join is bounded by the size of input relations, and the output size of each join is bounded by $O(\min(N \OUT, F))$, the total cost is bounded by $O(\min(N \OUT, F))$.
\end{proof}

\section{Sub-Graph Pattern Benchmark (SGPB) Details}
\label{App:Graph}

Table \ref{graph} presents the types of $9$ queries set for the SNAP dataset, characterizing the queries from $4$ dimensions.
\begin{itemize}
    \item \textbf{Shape: } The shape of a query.
    \item \textbf{Type: } Indicating the type of query.
    \item \textbf{Predicates: } The number of predicates included within a query.
    \item \textbf{Free-Connex: } Whether the query is a free-connex query.
\end{itemize}

\begin{table}[H]
\centering
\caption{SNAP-Queries}
\resizebox{0.45\columnwidth}{!}{ 
\begin{tabular}{l|ccccc}
\toprule
\textbf{Query} & \textbf{Shape} & \textbf{Type} & \textbf{Predicates} & \textbf{Free-Connex}\\
\midrule
q1a & line-3   & Full Enumerate & $1$  & Yes\\
\midrule
q1b & line-3   & Aggregation & $0$  & Yes\\
\midrule
q1c & line-3   & Projection & $0$  & Yes\\
\midrule
q2a & dumbbell   & Full Enumerate & $1$  & Yes\\
\midrule
q2b & dumbbell   & Aggregation & $0$  & Yes\\
\midrule
q3a & line-3   & Full Enumerate & $1$  & Yes\\
\midrule
q3b & line-3   & Aggregation & $0$  & Yes\\
\midrule
q3c & line-3   & Projection & $0$  & Yes\\
\midrule
q4a & line-5   & Projection     & $0$  & Yes \\
\midrule
q4b & line-5   & Aggregation     & $0$  & Yes \\
\midrule
q5a & line-5   & Projection     & $0$  & Yes \\
\midrule
q5b & line-5   & Aggregation     & $0$  & Yes \\
\midrule
q6 & line-3   & Projection    & $0$  & No \\
\midrule
q7 & line-4   & Aggregation    & $0$  & No \\
\midrule
q8 & line-4  & Aggregation    & $0$  & No \\
\midrule
q9 & line-4  & Aggregation     & $0$  & No \\
\bottomrule
\end{tabular}
}
\label{graph}
\end{table}

 \end{document}